\documentclass[10pt]{article}
\usepackage{amssymb}
\usepackage{amstext}
\newcommand*\de{\mathrm{d}}
\newcommand*\De{\mathrm{D}} 
\renewcommand*\epsilon{\varepsilon}
\renewcommand*\phi{\varphi}
\renewcommand*\theta{\vartheta}
\begin{document}

\title{\bf \large Noether's theorem, the stress-energy tensor and 
Hamiltonian constraints} 

\author{\small M. Leclerc \\ \small Section of Astrophysics and Astronomy, 
Department of Physics, \\ \small University of Athens, Greece}  
\date{\small August 3, 2006}
\maketitle 
\begin{abstract}
Noether's theorem  
is reviewed  with a particular focus on an intermediate 
step between global and local  gauge and coordinate 
transformations, namely linear transformations. We rederive the well 
known result that global symmetry leads to charge conservation (Noether's 
first theorem), 
and show that linear symmetry allows for the current to be expressed 
as a four divergence.  Local symmetry leads to 
identical conservation of the current and allows for the expression 
of the charge as two dimensional surface integral (Noether's 
second theorem). In the context of 
coordinate transformations, an additional step  (Poincar\'e 
symmetry) is of physical interest and leads to the definition of 
the symmetric Belinfante stress-energy tensor, 
which is then shown to be identically zero in generally covariant 
first order theories. The intermediate step of linear symmetry 
turns out to be important in general relativity when the 
customary first order Lagrangian is 
used, which is covariant only under affine transformations. 
In addition,  we derive explicitely the  
canonical stress-energy tensor in second order theories in its 
identically conserved form. Finally, we analyze the relations between 
the generators of local transformations, the corresponding 
currents and the Hamiltonian constraints. 
\end{abstract}

\section{Introduction}
Our motivation originates in the following question: Why is it that in 
gravity theory, the stress-energy (pseudo)\-tensors usually presented in 
literature are always in a form such that they are identically divergence 
free, while this is not the case for special relativistic theories? 
One could try to attribute this to the special role played by the 
gravitational field, which is supposed to define the geometry of spacetime 
 itself. But on the other hand, one can also take a less geometric point 
of view and forget about metrical structures and geometry in general. 
We are then left with a set of points $x^i$ (the spacetime manifold) 
and fields defined on this set, 
one of which being the gravitational field $g_{ik}$. From this viewpoint, 
there is no particular reason why the stress-energy tensor for the 
gravitational 
field should have any essentially different properties from that of any 
other field. 

The answer to the above question is given by Noether's second 
theorem. 
While global symmetry of a Lagrangian theory leads to 
a conservation law (first Noether theorem), invariance of the theory under 
the same, but localized  symmetry leads to identical conservation 
of the same current, in the sense that the charge can be written in the form 
of a surface integral. The exact formulation of the theorem can be found 
in the corresponding literature, in particular in Noether's original  
article \cite{noether}. 
In this paper, we will review what it  means concretely in the case of 
gauge and in particular of coordinate transformations. The answer to our 
initial question is provided by the fact that gravitational theories are 
 covariant under general coordinate transformations, 
while special relativistic theories are only Poincar\'e 
invariant. 

Since the results presented in this paper were already known to  
Noether herself, there is no need to provide a large reference list. 
Our aim is, on one hand, to present the results in a mathematically simple 
form (as opposed to fiber bundle descriptions) 
for the cases of physical interest, and, on the other hand, 
 to highlight the relevance of 
the different degrees of {\it locality} of the symmetry.  

Our main focus lies on the stress-energy tensor and the 
conservation of energy and momentum,  but in order to illustrate 
the procedure, and in particular the role of the \textit{intermediate} step 
between global and local transformations, we start with a simple example of 
an internal gauge symmetry. In the first five sections of this paper, 
we assume that the Lagrangian $\mathcal L$ does not depend on 
second and higher derivatives of the fields. We refer to such theories 
as {\it first order} theories. Second order theories will be considered 
in section 6. 

The paper is structured as follows. In the remaining part 
of the Introduction, we derive Noether's theorem for first order 
non-abelian gauge theory. Then, in sections 2 to 4, we turn to 
coordinate transformations and the stress-energy tensor. We start with 
global translations and Poincar\'e symmetry in section 2, go over to 
affine symmetry  
in section 3, and finally to general covariance in section 4, 
where, in each step, the consequences of the symmetry on the 
conservation law for the canonical as well as for the Belinfante 
stress-energy tensor are discussed. In section 5, 
we apply those results to concrete theories, 
in particular to general relativity and to Einstein-Cartan theory.     
In section 6, we generalize the formalism to include theories 
based on Lagrangians containing second derivatives of the fields. Further, 
in section 7, we briefly analyze the relations between the  
Belinfante and the Hilbert (or metric) stress-energy of the matter fields.  
Finally, in section 8, we analyze the relations between the generators 
of gauge and coordinate transformations, the corresponding currents and 
the first class Hamiltonian constraints of the theory.

We begin by considering 
 a Lagrangian $\mathcal L$ depending on a matter field $\psi$ 
and a gauge field $A^{\alpha}_i$, and assume  that $\mathcal L$ is invariant 
under the following  transformation   
\begin{equation} \label{1}
\delta A^{\alpha}_i = \epsilon^{\alpha}_{,i} + c_{\beta\gamma}^{\ \ \alpha} 
A^{\gamma}_i \epsilon^{\beta}, \ \ \ \delta \psi = - i \epsilon^{\alpha} 
\sigma_{\alpha} \psi. 
\end{equation}
The variation  of $\mathcal L$ then 
reads\footnote{In order to shorten our expressions, 
we omit the terms in $\bar \psi$, which are similar to those in $\psi$.}  
\begin{displaymath} 
\delta \mathcal L  = 0 = 
\frac{\partial \mathcal L}{\partial \psi} \delta \psi +
\frac{\partial \mathcal L}{\partial \psi_{,i}} \delta \psi_{,i}+
\frac{\partial \mathcal L}{\partial A^{\alpha}_{k}} \delta A^{\alpha}_k+
\frac{\partial \mathcal L}{\partial A^{\alpha}_{k,i}} \delta
A^{\alpha}_{k,i}. 
\end{displaymath}
Taking into account the field equations 
$\partial \mathcal L/\partial \psi  =  
(\partial \mathcal L/\partial \psi_{,i} )_{,i}$ and 
$\partial \mathcal L/\partial A^{\alpha}_k  =  
(\partial \mathcal L/\partial A^{\alpha}_{k,i} )_{,i}$, we find 
\begin{displaymath}
0=\left[-i \frac{\partial \mathcal L}{\partial \psi_{,i}} \sigma_{\alpha} \psi 
+ \frac{\partial \mathcal L}{\partial A^{\beta}_{k,i}} 
c_{\alpha\gamma}^{\ \  \beta} A^{\gamma}_k\right]_{,i}\epsilon^{\alpha} 
+
\left[-i \frac{\partial \mathcal L}{\partial \psi_{,i}} \sigma_{\alpha} \psi 
+ \frac{\partial \mathcal L}{\partial A^{\beta}_{k,i}} 
c_{\alpha\gamma}^{\ \  \beta} A^{\gamma}_k  + (\frac{\partial \mathcal L}
{\partial A^{\alpha}_{i,k}})_{,k}\right]  \epsilon^{\alpha}_{,i} 
+ \frac{\partial \mathcal L}{\partial A^{\alpha}_{k,i}}
  \epsilon^{\alpha}_{,i,k}. 
\end{displaymath}
Let us define the current by $J^{  i}_{\alpha} =  
-i \frac{\partial \mathcal L}{\partial \psi_{,i}} \sigma_{\alpha} \psi 
+ \frac{\partial \mathcal L}{\partial A^{\beta}_{k,i}} 
c_{\alpha\gamma}^{\ \  \beta} A^{\gamma}_k$.  
In a first step, we assume invariance of $\mathcal L$ under  
global transformations, 
i.e., $\epsilon^{\alpha}_{,i} = 0$. We then find the well known 
conservation law  
\begin{equation} \label{2}
J^{ i}_{\alpha \ ,i}  = 0. 
\end{equation}
Next, we assume that the Lagrangian is  invariant (in addition) 
under linear transformations, i.e., 
$\epsilon^{\alpha} = \epsilon^{\alpha}(x)$, 
with $\epsilon^{\alpha}_{,i,k} = 0$. 
This leads to the relation  
\begin{equation}\label{3}
J^i_{\alpha} =- (\frac{\partial \mathcal L}{\partial A^{\alpha}_{i,k}})_{,k}, 
\end{equation}
which expresses the current in the form of a divergence. In a last step, we 
assume local gauge invariance (i.e., general $\epsilon^{\alpha}(x)$), which 
gives us a third relation  
\begin{equation} \label{4}  
\frac{\partial \mathcal L}{\partial A^{\alpha}_{(i,k)}} =  0. 
\end{equation}
To summarize, global invariance leads to the conservation 
law (\ref{2}) for the current (and thus to charge conservation). 
Linear invariance allows us to write the current in 
the form (\ref{3}), and finally, local symmetry leads to (\ref{4}), which 
tells us that the expression (\ref{3}) is the divergence of an antisymmetric 
quantity, which as such has an identically vanishing divergence. Thus, 
in a locally invariant theory, the current can be written in a form 
such that it is identically conserved. In particular, this means that the 
charge (integration over a spacelike hypersurface) 
\begin{equation} \label{5}
Q_{\alpha}= \int J^i_{\alpha} \de \sigma_i = \int J^0_{\alpha} \de^3 x 
\end{equation} 
can be written in the form (greek indices from the middle of the alphabet, 
$\mu,\nu\dots$ refer to the spatial components and run from 1 to 3)
\begin{equation} \label{6}
Q_{\alpha} = - \oint \frac{\partial \mathcal L}
{\partial A^{\alpha}_{i,k}} \de \sigma_{ik}
= - \oint  \frac{\partial \mathcal L}
{\partial A^{\alpha}_{0,\mu}} \de^2 \sigma_{\mu}, 
\end{equation}    
i.e.,  in the form of a two dimensional surface integral. Note that this would 
not be possible without the relation (\ref{4}). In the abelian case, 
equation (\ref{6}) is nothing but the well known expression of the charge 
in form of an integral of $\vec E$ over a closed surface.

Similar results will be derived in the next sections for coordinate 
transformations. 

\section{Translations and Poincar\'e transformations} 

We consider now a generic Lagrangian $\mathcal L = \mathcal L(\phi,
\phi_{,i})$,  where $\phi$ denotes collectively all the fields 
(including the gravitational field, whenever present). Under a 
global coordinate transformation $x^i \rightarrow x^i - \xi^i$, with 
constant $\xi^i$, 
all fields (scalar, vector, tensor, spinor) transform as 
\begin{equation} \label{7} 
\delta \phi = \phi_{,i} \xi^i, 
\end{equation}
where $\delta \phi$ denotes the change of $\phi$ at the point with 
the same coordinates, i.e., 
$\delta \phi(x) = \phi'(x) - \phi(x)$. If we assume invariance of 
the action under global coordinate translations, the Lagrangian 
must transform as scalar and we find 
\begin{displaymath}
\delta \mathcal L = \mathcal L_{,i} \xi^i = 
\frac{\partial \mathcal L}{\partial \phi} \delta \phi + 
\frac{\partial \mathcal L}{\partial \phi_{,i}} \delta \phi_{,i}  
\end{displaymath}
which leads to the conservation law 
\begin{equation}\label{8} 
\tau^i_{\ k,i}  = 0, 
\end{equation}
with the canonical stress-energy tensor 
\begin{equation}\label{9} 
\tau^i_{\ k} \equiv \frac{\partial \mathcal L}{\partial \phi_{,i}} \phi_{,k} 
- \delta^i_k  \mathcal L.  
\end{equation}
Note that, although $\mathcal L$ will, in generally covariant theories, 
be a scalar density rather than a scalar, we will nevertheless stick to 
the definition (\ref{9}) (thus including the factor $\sqrt{-g}$ into
$\tau^i_{\  k}$).  

Thus, our first result is that global translational symmetry 
leads to a conserved energy-stress tensor. It is not hard to see that 
every Lagrangian that does not depend explicitely on $x^i$ possesses that 
symmetry. 

Before we turn to linear transformations, we consider Poincar\'e
transformations, which are of particular physical interest. The fields 
now transform as 
\begin{equation}\label{10}
\delta \phi = \xi^i \phi_{,i} + \frac{1}{2} \epsilon^{ik} 
(S \phi)_{ki}, 
\end{equation}
with $\xi^i = a^i + \epsilon^i_{\ k} x^k $ 
(constant $a^i$ and $\epsilon^i_{\  k}$) and $\epsilon^{ik} = \epsilon^i_{\ l}
\eta^{lk} = - \epsilon^{ki}$. By $(S \phi)_{ki}$, we denote the 
action of the Lorentz group on the field in question. Explicitely, we have 
\begin{eqnarray}\label{11}
(S \phi)_{ki} = 0 &\ \ \ &\text{scalar} \\\label{12}
(S A_l)_{ki} = \eta_{kl} A_i - \eta_{il} A_k &\ \ \ & \text{vector} \\
(S h_{lm})_{ki} = \eta_{kl} h_{im} - \eta_{il} h_{km} + \eta_{km} h_{il}
- \eta_{im} h_{kl} &\ \ \ & \text{tensor},    \label{13}
\end{eqnarray}
and similar for contravariant or mixed tensors $A^i, h^{ik}, h^i_{\ k}$. 
Note that, since we 
are ultimately interested in generally covariant theories, we have to assume 
that spinors (spin 1/2) are described by fields transforming 
as scalars under Lorentz transformations\footnote{Spin 3/2 fields 
$\psi_l$ carry an additional vector index  and transform according to 
(\ref{12}).} 
(since else, the generalization to 
 general linear and diffeomorphism transformations would not be possible 
with finite dimensional representations). Such fields differ from true 
scalar fields (spin 0) by their behavior under  local Lorentz gauge 
transformation, unrelated to the coordinate transformations we consider here. 

Assuming Poincar\'e invariance of the action, and thus $\delta \mathcal L = 
\mathcal L_{,i} \xi^i$, we find after some simple manipulations, apart from 
(\ref{8}), the additional relation  
\begin{displaymath}
(\tau^{ik} +  \frac{1}{2} S^{mik}_{\ \ \ ,m}) \epsilon_{ik} = 0, 
\end{displaymath} 
where $\tau^{ik} = \eta^{kl} \tau^i_{\ l}$ and 
\begin{equation}\label{14}
S^{mik} = \frac{\partial \mathcal L}{\partial \phi_{,m}} 
(S \phi)^{ik}.  
\end{equation}
Thus, since $\epsilon^{ik}$ is arbitrary (but antisymmetric), we get 
\begin{equation} \label{15}
\tau^{[ik]} +  \frac{1}{2} S^{mik}_{\ \ \ ,m} = 0.  
\end{equation}
This result can be compared with (\ref{3}), but it is less strong since 
it determines only the form of the antisymmetric part or $\tau^{ik}$. 
However, (\ref{15}) can be used for a different  purpose, namely the 
symmetrization of $\tau^{ik}$. Indeed, we can define the so-called 
Belinfante tensor (see \cite{belinfante} and \cite{jackiw})  
\begin{equation} \label{16}
T^{ik} \equiv \tau^{ik} + \frac{1}{2} [S^{ikm} - S^{mki} - S^{kmi}]_{,m}, 
\end{equation} 
which obviously satisfies $T^{ik}_{\ \ ,i} = 0$, since the expression in 
brackets is antisymmetric in $im$. In other words, 
$T^{ik}$ differs from $\tau^{ik}$ only 
by a so-called relocalization term of the form $C^{imk}_{\ \ \ ,m}$, 
with $C^{imk} = - C^{mik}$,  
i.e., a term whose 
 divergence $C^{imk}_{\ \ ,m,i}$ 
vanishes identically and which leads only to two dimensional 
surface terms in the field  momentum. Also, using (\ref{15}), it follows 
immediately that we have 
\begin{equation} \label{17} 
T^{ik} = T^{ki}.  
\end{equation}
Therefore, we can also define a spin current density in the form 
$\sigma^{lki} = T^{il}x^k - T^{ik}x^l$, satisfying $\sigma^{lki}_{\ \ \ ,i} 
= 0$, which is not possible with the asymmetric tensor $\tau^{ik}$.  

We will further discuss the Belinfante tensor later on. 

\section{Linear transformations} 

We now replace the Lorentz group by general linear transformations, 
considering Lagrangians with an  affine symmetry. The fields 
now transform as 
\begin{equation}\label{18}
\delta \phi = \xi^i \phi_{,i} + \frac{1}{2} \epsilon^{i}_{\ k} 
(\sigma \phi)^k_{\ i}, 
\end{equation}
with $\xi^i = a^i + \epsilon^i_{\ k} x^k$, with general (constant) 
$\epsilon^i_{\ k}$, and where the action of the general linear group 
on the fields is given by 
\begin{eqnarray}\label{19}
(\sigma \phi)^{k}_{\ i} = 0 &\ \ \ &\text{scalar} \\\label{20}
(\sigma A_l)^k_{\ i} = 2 \delta^k_l A_i &\ \ \ & \text{vector} \\
(\sigma h_{lm})^k_{\ i} 
= 2 (\delta^k_l  h_{im} + \delta^k_m h_{il})
 &\ \ \ & \text{tensor}.     \label{21}
\end{eqnarray}
 Let us also define 
\begin{equation}\label{22}
L^{mi}_{\ \ k} = \frac{\partial \mathcal L}{\partial \phi_{,m}} 
(\sigma \phi)^i_{\ k}.  
\end{equation}
Assuming invariance of the action means 
that $\mathcal L$ transforms as scalar density, and thus, $\delta 
\mathcal L = (\mathcal L \xi^i)_{,i} = \mathcal L_{,i} \xi^i 
+ \mathcal L \xi^i_{\ ,i}$. For the rest, the argument goes just as 
in the case of the Lorentz symmetry, with the only difference that we 
end up with 16 (instead of 6) independent equations. The result is 
\begin{equation} \label{23}
\tau^i_{\ k} = - \frac{1}{2} L^{mi}_{\ \  k,m} 
\end{equation} 
This is now in full analogy to equation (\ref{3}), in the sense that 
invariance under  linear transformations (together with the global 
translations) dictates the form of the conserved current $\tau^i_{\ k}$, 
which is again in the form of a divergence. 

In contrast to the 
Lorentz case, where the Minkowski metric $\eta_{ik}$ had been introduced 
by hand, 
no metric is needed for the above arguments. However, in order 
to make contact with  the results obtained for the Poincar\'e group, 
let us define 
\begin{equation} \label{24}   
L^{mik} = \eta^{kl} L^{mi}_{\ \ l}. 
\end{equation}
Then, we have obviously $S^{mik} = L^{m[ik]}$, and  from the expressions 
(\ref{16}) and (\ref{23}), we can derive the following expression for 
the symmetric Belinfante tensor
\begin{equation} \label{25}
T^{ik} = - \frac{1}{4}[L^{mik} - L^{ikm} + L^{imk} + L^{mki} + 
L^{kmi} - L^{kim}]_{,m} 
\end{equation}
A few remarks are in order at this point. First, if $\tau^i_{\ k} $ 
has the form of a relocalization term, i.e., if $L^{mi}_{\ \ k}$ in (\ref{23}) 
is antisymmetric in $im$, then it follows from (\ref{25}) that 
$T^{ik}  = 0$ identically. Second, it should be noted that (\ref{23}) and 
(\ref{25}) are very strong relations. Indeed, for a scalar field, e.g., 
they lead (in view of (\ref{19})) immediately to $\tau^i_{\ k} = 0$. 
The reason, however, is also obvious: It is not possible to construct a 
theory  that is invariant under general linear transformations only with 
scalar fields. For gauge fields (i.e., vector fields), we find 
\begin{equation}\label{26} 
\tau^i_{\ k} 
= - (\frac{\partial \mathcal L}{\partial A^{\alpha}_{i,m}}
A^{\alpha}_k)_{,m}. 
\end{equation}
We see that $\tau^i_{\ k,i} = 0$ is identically satisfied if 
$\partial \mathcal L/ \partial A^{\alpha}_{i,m}$ is antisymmetric in $im$. 
This is 
indeed the case in the physically relevant situations, where the derivatives
of the gauge 
fields enter only via the Yang-Mills tensor $F^{\alpha}_{ik}$. However, there 
could be exceptions, in particular concerning derivative couplings, e.g., 
of the tetrad field $e^a_i$ in gravitational theories without independent 
connection. This turns out not to be the case though, as we will 
now show in general. 

\section{General covariance} 

From the explicit form (\ref{23}), we see that only vector and tensor fields 
give explicit contributions to $\tau^i_{\ k}$. Thus, the question is 
what kind of restriction do we get on the form of those contributions 
from general covariance of the theory? Well, this is not difficult to 
find out. Under general coordinate transformations $x^i \rightarrow x^i - 
\xi^i$, the vector and tensor fields transform as\footnote{Those relations 
are simply the Lie derivatives of the corresponding fields.} 
\begin{eqnarray} \label{27}
\delta A_l &=& \xi^i A_{l,i} + \xi^i_{\ ,l} A_i, \\
\delta g_{lm} &=& \xi^i g_{lm,i} + \xi^i_{\ ,l} g_{im} + \xi^i_{\ ,m} g_{li}, 
\end{eqnarray}
while for the Lagrangian, we must have $\delta \mathcal L 
= (\xi^i \mathcal L)_{,i}$. Similar as in the the case of 
gauge symmetry considered in the Introduction, 
it turns out that it is actually enough to consider transformations of 
second order in $x^i$, i.e., $\xi^i = \epsilon^i_{\ kl} x^k x^l$.  
In any case, for the physically important case, namely the vector fields, 
we find 
\begin{displaymath} 
\frac{\partial \mathcal L}{\partial A^{\alpha}_{(i,m)}}A_k = 0, 
\end{displaymath}
or, in other words, 
\begin{equation} \label{29} 
L^{mi}_{\ \ k} = - L^{im}_{\ \ k}. 
\end{equation} 
The same result is found for the case of  tensor fields, but this is not 
really of interest, since there is no generally covariant action for a tensor 
field containing only first derivatives. In any case, we will give a general 
proof (for theories containing up to second derivatives)  in section 6. 
Equation (\ref{29}) is the 
analogue of equation (\ref{4}) and simply means that the canonical 
stress-energy tensor (\ref{23}) is identically conserved. In particular, 
the momentum vector 
\begin{displaymath}  
P_k = \int \tau^i_{\ k} \de \sigma_i = \int \tau^0_{\ k} \de^3 x 
\end{displaymath}
can be written as a two dimensional surface integral 
\begin{equation}\label{29a}
P_k =  \frac{1}{2} \oint L^{mi}_{\ \ k} \de \sigma_{mi} = - \frac{1}{2} \oint 
L^{\mu 0}_{\ \ k} \de^2 \sigma_{\mu}. 
\end{equation}
Moreover, as 
mentioned before, a consequence of (\ref{29})  is that the Belinfante tensor 
vanishes identically (a result that has been found in \cite{gotay}). 

As a result, we have shown that in generally covariant theories, the 
canonical stress-energy tensor can be written in the form of a relocalization 
term, and thus, the momentum vector in form of a two dimensional surface 
integral (see also \cite{sardan}), 
while the Belinfante tensor is identically zero. This holds for 
theories based on Lagrangians containing only first derivatives of the fields. 

\section{Applications} 

\subsection{Special relativity} 

The physically relevant theories that are generally covariant contain  
necessarily the gravitational field. Nevertheless, in order to 
illustrate our formalism, let us consider the following Lagrangian which is 
independent of any metrical background 
\begin{displaymath}
\mathcal L = \epsilon^{iklm}A_{i,k} A_{l,m}.  
\end{displaymath}
This Lagrangian is  a total divergence, and therefore, the field 
equations are identically satisfied for any $A_i$. 
The stress-energy tensor, according to (\ref{9}) is of the form 
\begin{displaymath}
\tau^i_{\ k} = 2 \epsilon^{liqp} A_{q,p} A_{l,k}
- \delta^i_k \epsilon^{rsqp}A_{rs} A_{pq}, 
\end{displaymath}
while from (\ref{23}), we find 
\begin{displaymath}
\tau^i_{\ k} =  2 \epsilon^{liqp} A_{q,p} A_{k,l},  
\end{displaymath}
which is obviously identically conserved. 
Also, the difference of both expressions, $\epsilon^{imqp}F_{km} F_{pq} 
- \frac{1}{4} \delta^i_k \epsilon^{lmpq} F_{lm}F_{pq}$ is easily shown to 
be zero. A similar calculation shows that  the Belinfante tensor $T^{ik}$
 is zero. 

\subsection{Einstein-Cartan theory} 

Einstein-Cartan theory is based on a tetrad field 
$e^a_i $ and a Lorentz connection $\Gamma^{ab}_{\ \ i}$. 
(Latin indices from the beginning of the alphabet $a,b,c\dots$ refer to 
  tangent space, with Minkowski metric $\eta_{ab} = diag(1,-1,-1,-1)$.)
For 
completeness, we consider the Einstein-Cartan-Dirac-Maxwell Lagrangian 
\begin{equation}\label{31} 
\mathcal L = - \frac{e}{2}  e^i_a e^k_b R^{ab}_{\ \ ik} -  
\frac{e}{4} F^{ik}F_{ik} + e [\frac{i}{2}(\bar \psi \gamma^m \De_m \psi - 
\De_m \bar \psi \gamma^m \psi) - m \bar \psi \psi], 
\end{equation}
with $e = \det e^a_i,\ R^{ab}_{\ \ ik} = \Gamma^{ab}_{\ \ k,i} 
- \Gamma^{ab}_{\ \ i,k} 
+ \Gamma^a_{\ ci} \Gamma^{cb}_{\ \ k} - \Gamma^a_{\ ck} \Gamma^{cb}_{\ \ i},\ 
 F_{ik} = A_{k,i} - A_{i,k}$ and $ \De_i \psi = 
\partial_i \psi - \frac{i}{4} \Gamma^{ab}_{\ \ i} \sigma_{ab} \psi
- i q A_i \psi$. From (\ref{23}) and (\ref{20}) , we  find 
\begin{equation} \label{32} 
\tau^i_{\ k} = - (e e^i_a e^m_b \Gamma^{ab}_{\ \ k})_{,m} 
- (e F^{im} A_k)_{,m}. 
\end{equation}
Obviously $\tau^i_{\ k,i} = 0$ identically, and $T^{ik} = 0$, according to 
our general theorem. The expression is easily generalized to non-abelian gauge 
fields $A_i^{\alpha}$.  

In order to evaluate the momentum $P_k$ in (\ref{29a}), it is useful to 
make a few general considerations. First, it is clear that, in the 
presence of radiation, we will, in general, not find a finite expression 
when the surface of integration is extended to spatial infinity. On the 
other hand, if we assume that the leading order contributions have the 
behavior $A_i = \mathcal O(\frac{1}{r})$, and thus $F_{ik} = \mathcal 
O(\frac{1}{r^2})$, then it is clear that the second term in (\ref{32}) 
does not contribute to $P_k$\footnote{For the boundary conditions 
necessary to obtain a finite expression for the conserved charge, see 
also \cite{jackiw}.}.
The same holds for non-abelian gauge 
fields, and in particular for $\Gamma^{ab}_{\ \ i}$ itself, 
if we include in  $\mathcal L$ terms quadratic in the curvature 
$R^{ab}_{\ \ ik}$. The only exception to this is the tetrad field, 
which can be assumed to behave like 
$e^a_i = \delta^a_i + \mathcal O(\frac{1}{r})$. Thus, quite generally, 
the only contributions that enter explicitely the momentum $P_k$ 
are those stemming from the Einstein-Cartan Lagrangian (first term in 
(\ref{32})), and eventually from additional terms in the Lagrangian, 
quadratic in the torsion tensor $T^a_{ik} = e^a_{k,i} + \Gamma^a_{\ bi} e^b_k 
- e^a_{i,k} - \Gamma^a_{\ bk} e^b_i$. Such terms, however, will in general 
also modify the Newtonian limit of the gravitational theory. 

Note that for the same reasons, relocalization terms are usually considered 
not to modify the momentum $P_k$, and are therefore used as a tool to 
modify the stress-energy tensor according to our will, without changing 
the physically important quantity, which is $P_k$. 
If this were generally true, however, then in
generally covariant theories, the 
momentum would always be zero, since we can always write the stress-energy 
tensor in the form of a relocalization term. It is therefore very important 
that, as we have argued, the linear Einstein-Cartan 
term $e R$ provides an exception to this. On the other hand, for the same 
reason, it is quite a questionable procedure, in the framework of those
theories,   to modify the stress-energy 
tensor by relocalization terms, and, e.g., define the Belinfante tensor. 
Those procedures were invented in special relativistic theory exactly because  
they are supposed not to modify the momentum, and not simply to overcome our 
dissatisfaction with an asymmetric tensor. After all, as we have shown in the
introduction, the electric current density too can be written in the form of 
a relocalization term, i.e., $j^i = C^{ik}_{\ \ ,k}$, with antisymmetric 
$C^{ik}$. However, would anyone ever come up with the idea to add 
an additional relocalization term  
because he does not like the specific 
form of $j^i$? And even if he did, he would certainly take care that 
at least the charge itself is not modified by this procedure. 

Let us return to the expression (\ref{32}). Recall that the connection 
can be split into a torsionless part and the contortion tensor 
$\Gamma^{ab}_{\ \ i} = \hat \Gamma^{ab}_{\ \ i} + K^{ab}_{\ \ i}$, 
where, in Einstein-Cartan theory, the contortion tensor is directly 
expressed in terms of the spin of the spinor field. In the absence of 
spinor fields, we have $K^{ab}_{\ \ i}= 0 $, and we can evaluate 
$\hat \Gamma^{ab}_{\ \ i}$ using $e^a_{i,k} + \hat \Gamma^a_{\ bk} e^b_i 
= e^a_l \Gamma^l_{ik}$, where $\Gamma^l_{ik}$ are the Christoffel symbols. 
For the Schwarzschild metric  
$\de s^2 = (1-M/r)\de t^2 - (1- M/r)^{-1} \de r^2 - r^2 
(\de \theta^2 + \sin^2 \theta \de \phi^2)$ 
and using a  diagonal tetrad, we  find $P_k =  2 \pi M \delta^0_k$,   
i.e., the energy is just one half of the mass, $P_0 = m/2$  ($ M = 2 k  m$, 
with $8 \pi k = 
1 $ in our units). The same result was found in the preprint version of 
\cite{jackiw}. Note that the same result emerges from the Reisner-Nordstroem 
solution (since only the leading order term of $g_{00} = -1/g_{rr}= 1- 
M/r + q^2/r^2$ contributes to the surface integrals at spatial infinity).   

Since $\tau^i_{\ k}$ is asymmetric, and $T^{ik} = 0$ identically, we cannot 
formulate a conservation law for the angular momentum based on the 
stress-energy tensor. However, we can exploit the Lorentz gauge 
symmetry of the theory, i.e., the invariance of the Lagrangian under 
\begin{equation} \label{33} 
\delta e^a_i = \epsilon^a_{\ b} e^b_i, \  \ \delta \Gamma^{ab}_{\ \ i} 
= - \De_i \epsilon^{ab}, \ \ \delta \psi = - \frac{i}{4} \epsilon^{ab}
\sigma_{ab} \psi,
\end{equation}
where $\epsilon^{ab} = - \epsilon^{ba}$ and 
$\De_i  \epsilon^{ab} = \epsilon^{ab}_{\ \ ,i} + \Gamma^a_{\ ci} 
\epsilon^{cb} +  \Gamma^b_{\ ci} \epsilon^{ac}$. Let us define the 
following current\footnote{For simplicity, we omit again the terms 
with $\bar \psi$.} 
\begin{equation} \label{34}
J_{ab}^{\ k} = 
\frac{1}{2} \frac{\partial \mathcal L}{\partial e^a_{i,k}} e_{bi} 
-\frac{1}{2} \frac{\partial \mathcal L}{\partial e^b_{i,k}} e_{ai} 
- \frac{i}{4} 
\frac{\partial \mathcal L}{\partial \psi_{,k}} \sigma_{ab} \psi -
\frac{\partial \mathcal L}{\partial \Gamma^{cb}_{\ \ i,k}}\Gamma^c_{\ ai} -
\frac{\partial \mathcal L}{\partial \Gamma^{ac}_{\ \ i,k}} 
\Gamma^{c}_{\ bi}, 
\end{equation}
which is antisymmetric in $ab$ and can be interpreted as angular momentum 
density current. Performing the same steps as in the introduction, 
we find from global symmetry ($\epsilon^{ab}$ constant)
\begin{equation} \label{35}
J_{ab\ ,k}^{\ \ k} = 0. 
\end{equation}
and from linear symmetry ($\epsilon^{ab}_{\ \ ,i,k} = 0$) the 
explicit form 
\begin{equation}\label{36} 
J_{ab}^{\ \ k}=  
(\frac{\partial \mathcal L}{\partial \Gamma^{ab}_{\ \ k,i}})_{,i},  
\end{equation}
and finally, from local symmetry that the expression in parentheses is 
antisymmetric in $ki$, such that the current is identically conserved 
and the corresponding charge can be put into the form of a two 
dimensional surface integral. Note that for the particular case of  
Einstein-Cartan theory, 
the first two terms in (\ref{34}) are automatically absent.  

Although we focused, in this section, mainly on Einstein-Cartan theory, 
it is obvious that the results are also valid for any candidate of 
Poincar\'e gauge theory, i.e., for any theory with an independent Lorentz 
connection and with a Lagrangian at most quadratic in curvature and 
torsion. We refer to \cite{blago}, where the Hamiltonian analysis of 
those theories has been carried out. 
The analysis is easily extended to metric affine theories 
(with an independent general linear connection, see \cite{hehl}), 
provided one finds a way to couple the spinor fields to the 
connection, see, e.g., \cite{leclerc3}.   

Finally, we should also mention that in some sense, the 
 Lorentz gauge group is more closely connected to the coordinate 
transformation group than conventional  internal gauge groups. In order 
to perform Wigner's classification of elementary particles, you need 
to analyze the behavior of the fields under both the diffeomorphism and 
the Lorentz group. The link is provided by the fact that in the flat limit, 
$e^a_i = \delta^a_i$, and the residual transformation freedom is then, 
apart from global coordinate translations, a  global Lorentz gauge 
rotation and a simultaneous 
global Lorentz coordinate transformation (with the same 
parameters), such that $e^a_i = \delta ^a_i$ remains unchanged. 
In other words, the Poincar\'e transformation group of special relativity 
(on which the particle classification is based) emerges in the flat limit 
as a combination of both the Lorentz gauge group and the diffeomorphism 
group. As a result, it is not unnatural to consider combinations of 
both groups right from the start. That is, instead of the conventional 
Lie derivatives $\delta \phi = \phi_{,i} \xi^i + \frac{1}{2} \xi^i_{\
 ,k}(\sigma \phi)^k_{\  i}$ related to pure diffeomorphism invariance, 
one can consider modified Lie derivatives under which the theory is 
still invariant,  as a result of the additional Lorentz gauge invariance.    
In this way, one can derive in quite a natural way 
alternative expressions for the conserved energy-momentum currents. 
The details of this procedure have been recently worked out in \cite{obukhov}.

\subsection{Tetrad gravity}

A more conservative way to incorporate spinor fields into general relativity 
is by simply replacing the metric tensor with the tetrad field, without 
introducing an independent connection. The gravitational field then 
couples to the spinor field via $\hat \De_i \psi = \partial_i \psi
- \frac{i}{4}\hat \Gamma^{ab}_{\ \ i}$, where $\hat \Gamma^{ab}_{\ \ i}$ 
is a function of the tetrad field and its derivatives that can be 
evaluated from $e^a_{i,k} + \hat \Gamma^a_{\ bk} e^b_i = e^a_l
\Gamma^l_{ik}$. The free gravitational Lagrangian is taken in the form  
\begin{equation} \label{37}
\mathcal L = - \frac{e}{2}[- \frac{1}{4} \tau^{ikl} \tau_{ikl} - \frac{1}{2} 
\tau^{ikl} \tau_{lki} + \frac{1}{2} \tau^m_{\ im} \tau^{li}_{\ \ l}],  
\end{equation}
 where $\tau^a_{\ ik} = e^a_{k,i} - e^a_{i,k}$. This Lagrangian coincides 
up to a surface term with the Hilbert-Einstein Lagrangian, 
but it has the additional 
feature that it does not contain second derivatives of the tetrad field, 
and it is generally covariant under coordinate transformations. Note also 
that $\mathcal L $ is invariant under 
(local) Lorentz rotations $\delta e^a_i = \epsilon^a_{\ b} e^b_i$. 
For this Lagrangian, we find 
\begin{equation} \label{38} 
L^{mi}_{\ \ k} = e [\tau_k^{\ mi} + \tau^{im}_{\ \ k}    
- \tau^{im}_{\ \ k} - \delta^i_k \tau^{lm}_{\ \ l} - \delta^m_k 
\tau^{li}_{\  l}], 
\end{equation}
which is antisymmetric in $im$, and thus leads an identically conserved  
$\tau^i_{\ k}$ and to $T^{ik} = 0$ for the Belinfante tensor. 

Note that, as a result of the derivative  couplings, in the presence of spinor 
fields (with the same Dirac Lagrangian as in (\ref{31}), where 
 $\De_i \psi$ is replaced by 
$\hat \De_i \psi$) we find an additional term in the form 
\begin{equation} \label{39} 
L^{mi(D)}_{\ \ k} = 
4 [\sigma_k^{\ mi} - \sigma_k^{\ im} + \sigma^{im}_{\ \ k}], 
\end{equation}  
where $\sigma_{ab}^{\ \ m} = \frac{\partial \mathcal L}{\partial
  \hat \Gamma^{ab}_{\ \ m}}$ is the so-called spin density of the spinor
  field. (This is not a conserved quantity, though.) 
Again, $L^{mi(D)}_{\ \ k} $ leads to a surface term. (There can also be 
additional terms, from gauge vector fields, in the form 
of the second term in (\ref{32}) which, as we have argued before, 
 do not contribute 
explicitely to the momentum.)

The theory  has a local Lorentz symmetry 
\begin{equation} \label{40} 
\delta e^a_i = \epsilon^a_{\ b} e^b_i, \  \ \delta \psi 
= - \frac{i}{4} \epsilon^{ab} \sigma_{ab} \psi,
\end{equation}
which could eventually be used to define an angular momentum 
 density for the system. 
 However, from the above symmetry, we find relations formally identical 
to  (\ref{34}), (\ref{35}) and (\ref{36}), without the terms involving 
$\Gamma^{ab}_{\ \ i}$. In particular therefore, from (\ref{36}), we 
find that the current vanishes identically. Thus, in this theory, it 
is not possible to formulate a conservation law for angular momentum, 
neither from the stress-energy tensor, nor using the 
local Lorentz symmetry. Note however that you can use the vanishing 
of the expression  (\ref{34}) in order to simplify the  
 stress-energy tensor obtained from (\ref{38}) and (\ref{39}), by 
observing that the second term in (\ref{34}) is again the spin density 
$\sigma_{ab}^{\ \ k}$. We then find $L^{m[ik]} = - 2\sigma^{ikm}$
 (for the total expression, (\ref{38}) plus (\ref{39})), and 
therefore, from (\ref{15}), $\tau^{[ik]} = \sigma^{ikm}_{\ \ \ ,m}$, 
showing that in the absence of spinor fields, $\tau^{ik}$ is automatically 
symmetric.

\subsection{General relativity}

Finally, we consider classical general relativity. Since we have 
restricted our formalism to Lagrangians that do not contain second 
derivatives of the fields, we start with the Lagrangian 
\begin{equation} \label{41} 
\mathcal L = -\frac{1}{2}\sqrt{-g}\ \left[ g^{ik}(\Gamma^l_{im} \Gamma^m_{kl}
-\Gamma^l_{ik} \Gamma^m_{lm})\right], 
\end{equation}
which is equivalent to the Hilbert-Einstein Lagrangian $-\frac{1}{2} \sqrt{-g}
R$. Apart from $g_{ik}$, there might be gauge vectors $A^{\alpha}_i$ 
and scalar fields $\phi$, which do not contribute explicitely to $P_k$ 
(and do not appear at all in the Belinfante tensor $T^{ik}$). 
We assume that there are no derivative couplings of $g_{ik}$ to any 
other fields, as is indeed the case with gauge fields and 
minimally coupled scalars. (Spinors cannot be described within a purely metric 
formalism.) After some algebra, we derive from (\ref{41}) the following 
expression 
\begin{eqnarray} \label{42} 
L^{mk}_{\ \ i} &=& \sqrt{-g}\ 
\left[g^{lm}_{\ \ ,l} \delta^k_i - g^{kl}_{\ \ ,l} 
\delta^m_i + g^{mq}_{\ \ ,l} g_{iq} g^{lk} - g^{kq}_{\ \ ,l} g_{iq} g^{lm}
\right]
\nonumber \\&&+\left[\sqrt{-g}\ (-2 g^{lm} \delta^k_i + g^{lk} \delta^m_i + 
g^{km} \delta^l_i)\right]_{,l}. 
\end{eqnarray}
The expression  in the first row is antisymmetric in $mk$ and is 
thus not only identically conserved, but moreover, it does not 
contribute to the Belinfante tensor $T^{ik}$, as we have established 
previously. 
The expression  in the second row, however, is not antisymmetric in $mk$.  
This is because $\mathcal L$ is not covariant  under general coordinate 
transformations. It is not hard to see that nevertheless, the divergence 
of this expression vanishes identically. The expression is in the form of what 
we will call in the next section a {\it second order } relocalization term.  
Also, since  $\mathcal L$ is covariant under linear 
transformations, the relation $\tau^i_{\ k} = - \frac{1}{2} L^{mi}_{\ \ k,m}$ 
is still valid. Note, however, that in order to evaluate
the total canonical stress-energy tensor, you have to add the contributions 
from the vector fields to $L^{mi}_{\ \ k}$. (Although they 
will in general not contribute to $P_k$ anyway.) 

Finally, since $L^{mk}_{\ \ i} $ is not 
antisymmetric in $mk$, the Belinfante  tensor $T^{ik}$ does not 
identically vanish, and indeed, we find from (\ref{25}) 
\begin{equation} \label{43}
T^{ik} = \frac{1}{2} \left[ \sqrt{-g}\ (\eta^{ik} g^{lm} - g^{im} \eta^{kl} 
- g^{km} \eta^{li} + g^{ik} \eta^{lm})\right]_{,l,m}. 
\end{equation}
This is exactly the form derived in \cite{jackiw} by the direct 
application of the Belinfante relation  (\ref{16}) to the canonical tensor 
(\ref{9}) (and a subsequent use of the Einstein field equations) 
and the tensor $T^{ik}$ is known as Papapetrou tensor. 
Note that matter fields (vector and scalar) do not 
contribute to $T^{ik}$, since they lead to antisymmetric (and zero) 
contributions in $L^{mi}_{\ \ k}$, and therefore $T^{ik}$ represents 
 the total stress-energy tensor of the system.

It is interesting to remark that $T^{ik}$, although apparently 
 not written in 
the form of a relocalization term (see, however, section 6), 
is  identically  conserved  
and it is also not hard to show that the momentum can again be written as 
a two dimensional surface integral, although two steps have to be performed 
to achieve this. 
(In the integral over $T^{0k}$ from (\ref{43}), 
split first one of the indices $l$ or $m$ 
into time and space components, and then the other one. The term with both 
$l=m=0$ does not contribute, while the remaining terms lead to surface 
integrals.) It turns out (see \cite{jackiw}) that 
the same Belinfante tensor emerges 
from the full (covariant) Hilbert-Einstein Lagrangian. (Not the same canonical 
tensor $\tau^i_{\ k}$, though.)
This shows that the statement 
given at the end of section 4 (i.e., that in generally covariant theories, 
the Belinfante tensor vanishes identically) does not generalize in its full 
extend to generally covariant 
theories with second 
derivatives in the Lagrangian. (However,  there is a generalization of that 
statement, in the sense that the stress-energy tensor $\tau^i_{\ k}$ is still 
identically conserved and the momentum can be 
expressed in terms of two dimensional surface integrals, similar as is the 
case with $T^{ik}$ here. We will show this in section 6.)   
    
We have evaluated the momentum emerging from $\tau^i_{\ k}$ 
and $T^{ik}$ and find $P_k =  4 \pi M \delta^0_k =  m \delta^0_k$ 
for the Schwarzschild metric in both cases, which is the double value of 
that found in Einstein-Cartan theory. This is in accordance with 
the results found in  
\cite{jackiw}, where in addition, the 
tensor $T^{ik}$ has been shown to be equivalent, 
as far as the leading order contributions are concerned, to the 
corresponding Landau-Lifshitz tensor and the Weinberg tensor. (Note that 
only the leading order contributions contribute to the surface integrals 
 at infinity anyway.) 

The Belinfante tensor (\ref{43}) being symmetric, we can formulate 
a conservation law for the  angular momentum density 
$T^{ik}x^m - T^{im} x^k$, which would not be possible from $\tau^i_{\ k}$ 
or from any other line of argumentation (as was the case in  
Einstein-Cartan theory). 

As a final remark, it is important to have in mind that the fact 
that the stress-energy tensor (both $\tau^i_{\ k}$ and $T^{ik}$) 
is identically conserved 
can not be derived merely from the symmetry of $\mathcal L$ under affine 
transformations, but it emerges here rather as an additional, unexpected 
result. Otherwise stated, not every Lagrangian covariant only under affine 
transformations leads 
to an identically conserved current. The specific Lagrangian 
considered here, however, has the additional property that it is 
equal, up to a four divergence, to a generally  covariant Lagrangian, albeit 
one of second order. As a result, under general coordinate transformations
 $\delta x^i = \xi^i$, 
the Lagrangian transforms as $\delta \mathcal L = (\xi^i \mathcal L)_{,i} 
+ \Phi^i_{\ ,i}$, i.e., it picks up  an additional four divergence. 
Quite obviously, such a 
transformation behavior is sufficient for the field equations to be 
generally covariant.

\subsection{Discussion}

Summarizing our results, we found that in Einstein-Cartan theory, 
the canonical tensor is identically conserved, while the Belinfante 
tensor vanishes. Nevertheless, it is possible to establish a 
conservation law for angular momentum, based on the local Lorentz 
gauge invariance of the theory. Thus, in a sense, there is no need 
for a symmetrized stress-energy tensor. The separation of the angular 
momentum from the stress-energy tensor (which, after all, is  
 the current corresponding to translational invariance) seems to be 
satisfying also from a logical point of view, if we recall that the spin 
structure of the spinor field was completely separated from the coordinate 
transformations (i.e., $\psi$ is treated as a scalar field), and was 
transferred to a tangent space with an inherent Lorentz symmetry. Thus, 
it should also be that same  Lorentz symmetry that is responsible for the 
angular momentum conservation. (The argument is not absolute, though, since 
after all, the angular momentum is not entirely given in terms of intrinsic 
spin.) 

On the other hand, in general relativity, we have a non-vanishing 
symmetric Belinfante tensor, and we can therefore formulate an  
angular momentum conservation law based on this tensor. Nevertheless, 
we see two problems with this procedure.  First, in the context of 
a generally covariant theory (or at least, covariant under the 
general linear group, in the first order theory), we arbitrarily pick  
 out a specific subgroup, namely the Lorentz group, and modify the 
canonical stress-energy tensor, which was already in the form of 
a relocalization term, by a relocalization term (of the same order of 
magnitude) such that the resulting tensor allows for the formulation 
of the conservation laws corresponding to that subgroup, namely 
angular momentum conservation. This, in our opinion, is not quite in the 
spirit of general relativity, which is based on the full diffeomorphism 
invariance, and has no inherent preferred subgroups. (Again, the argument 
is not absolute, since nevertheless, the Poincar\'e group  (or eventually 
the de Sitter group, in presence of a cosmological constant) emerges 
as groundstate symmetry of the theory, if the correct signature of the 
metric is assumed a priori.) Second, the metric framework of general 
relativity does not allow for the presence of spinor fields. Therefore, 
as soon as we deal with spinor fields, we have to go over to the 
tetrad formulation, which has therefore to be considered as more fundamental. 
However, as we have established,  the tetrad formulation 
leads to a vanishing Belinfante tensor, and there is no obvious way 
to define a conserved angular momentum current. Therefore, the apparent 
success of the Belinfante tensor in general relativity seems to be 
only a coincidence that does not generalize to the more fundamental 
formulation of the theory. And most importantly, the absence of a conserved 
angular momentum in the tetrad formulation of general relativity 
provides an argument in favor of Einstein-Cartan theory with 
independent Lorentz connection. 

Our opinion is therefore that the use of the Belinfante symmetrization 
procedure should be confined to the purpose it was initially designed for, 
namely to special relativistic theories,  where the Poincar\'e symmetry is 
an inherent ingredient right from the start, and where the integrated 
momentum is not influenced by the relocalization terms as a result of 
the asymptotic behavior of the fields in conventional theories.  
Both of those requirements  are violated in general relativity. 

Nevertheless, in the framework of general relativity, the tensor 
(\ref{43}) can certainly have its usefulness, if interpreted correctly and 
most importantly, if used consistently. 
It represents, after all, a conserved current and is,  
in this aspect,  no different from the Weinberg or the Landau-Lifshitz 
tensor (see \cite{jackiw}). But it is also clear that any expression 
of the form (\ref{43}), with $g^{lm}$ replaced with any other 
symmetric tensor, $\eta^{lm}$ replaced with any other symmetric tensor, 
and $\sqrt{-g}$ replaced with any other function, is identically  
conserved too, is symmetric too, and is also related to the 
canonical tensor by a relocalization term (of second order, see 
next section). Thus, modifying the canonical tensor by relocalization terms 
and requiring Poincar\'e invariance  does in no way fix 
the form of the stress-energy tensor.

\section{Second order field theory}

For the sake of completeness, we now extend our analysis to theories 
based on Lagrangians containing second derivatives of the fields (we 
refer to such theories as {\it second order}). 
The second Noether 
theorem tells us that again, invariance under a local symmetry 
leads to an identically conserved current. We will establish the explicit 
form of the current corresponding to general coordinate covariance. 

As can be expected, the analysis for second order theories 
contains one more step, namely 
global, linear, quadratic and finally general gauge or coordinate 
transformations. We start again with gauge theory as a warmup exercise, 
 but since we do not know of any  physically relevant second order gauge 
theory anyway, we  confine ourselves this time to the abelian theory. 
  
Thus, we consider a theory invariant under 
\begin{equation} \label{44}
\delta A_i = \epsilon_{,i}, \ \ \delta \psi = - i \epsilon \psi. 
\end{equation}
Having in mind that the field equations (for a generic field $\phi$) in 
second order theories are of the form 
\begin{equation} \label{45}
\frac{\partial \mathcal L}{\partial \phi} - 
(\frac{\partial \mathcal L}{\partial \phi_{,m}})_{,m}
+ (\frac{\partial \mathcal L}{\partial \phi_{,m,l}})_{,m,l} = 0, 
\end{equation}
and requiring $\delta \mathcal L = 0$ under (\ref{44}), first with 
constant $\epsilon$, then with linear $\epsilon$, then with 
second order $\epsilon$ (i.e., $\epsilon = \epsilon_{ik} x^i x^i$ with 
constant $\epsilon_{ik}$) and finally with general  $\epsilon(x)$, 
we arrive at  the following four equations
\begin{eqnarray} \label{46}
J^i_{\ ,i} &=& 0 \\  \label{47}
J^i &=& (\frac{\partial \mathcal L}{\partial \psi_{,i,k}} i \psi)_{,k} 
- (\frac{\partial \mathcal L}{\partial A_{i,k}})_{,k} 
- (\frac{\partial \mathcal L}{\partial A_{i,k,l}})_{,k,l}\\ \label{48}
0&=& \frac{\partial \mathcal L}{\partial \psi_{,i,k}} 
- \frac{\partial \mathcal L}{\partial A_{(i,k)}}\\\label{49}
0&=& \frac{\partial \mathcal L}{\partial A_{(i,k,l)}}, 
\end{eqnarray}
where the current has been  defined as  
\begin{equation} \label{50}
J^i \equiv - \frac{\partial \mathcal L}{\partial \psi_{,i}} i \psi 
- \frac{\partial \mathcal L}{\partial \psi_{,i,k}} i \psi_{,k} 
+ (\frac{\partial \mathcal L}{\partial \psi_{,i,k}})_{,k} i \psi.  
\end{equation}
Note that in (\ref{49}),  symmetrization  over the 
three indices is understood. The situation is essentially the same 
as in first order theories (see equations (\ref{2}), (\ref{3}), (\ref{4})). 
Global symmetry (\ref{46}) leads to current conservation, linear symmetry 
(\ref{47}) enables us to write the current in the form of a divergence, and 
finally, local symmetry (in this case, quadratic and cubic) shows that 
the current is identically conserved. More specifically, (\ref{48}) 
shows that the two first terms in (\ref{47}) are identically divergence 
free, while (\ref{49}) shows the same for the last term in (\ref{47}). 

Note that, if we take into account (\ref{48}) and (\ref{49}), 
then  the first two terms  in (\ref{47}) are in the form of the 
previously encountered relocalization terms, $C^{ik}_{\ \ ,k}$ with 
antisymmetric $C^{ik}$. It is interesting that the last term, although 
identically conserved, is not of the same form, but rather in the 
form $C^{ikl}_{\ \ ,k,l}$ with $C^{ikl}$ such that the totally symmetric 
part is zero, $C^{(ikl)} = 0$. It is not hard to show that such a 
term in the current $J^i$ leads again to a two dimensional surface term    
for the charge $\int J^0 \de^3 x$, namely ($\kappa, \lambda = 1,2,3$) 
\begin{displaymath}
\int C^{0kl}_{\ \ \ ,k,l} \de^3 x = 
\int C^{0k\lambda}_{\ \ \ ,k} \de^2 \sigma_{\lambda}
+ \int C^{0k0}_{\ \ \ ,0,k} \de^3 x = 
\int C^{0k\lambda}_{\ \ \ ,k} \de^2 \sigma_{\lambda}
+  \int C^{0\kappa 0}_{\ \ \ ,0}\ \de^2 \sigma_{\kappa},   
\end{displaymath} 
 where we use the fact that $C^{000} = 0$. Therefore, we will refer to 
such terms as {\it second order relocalization terms}\footnote{The 
quantities $C^{ik}$ and $C^{ikl}$ are also referred to as superpotentials.}. 
Note that 
the stress-energy tensor (\ref{43}) consists of such a term. 

We now turn to coordinate transformations in second order 
theories (see \cite{goldberg}). 
Thus, we consider again transformations of the form 
\begin{equation} \label{51}
\delta \phi = \xi^i \phi_{,i} + \frac{1}{2} \xi^i_{\ ,k} 
(\sigma \phi)^k_{\  i}, 
\end{equation}  
where the expressions for $(\sigma \phi)^k_{\ i}$ are given in
(\ref{19}), (\ref{20}) and (\ref{21}). Then, we require invariance of 
the action, i.e., $\delta \mathcal L = (\mathcal L \xi^i)_{,i}$ 
and derive the corresponding conservation laws. The manipulations 
are simple and we give only the results here. 

From invariance under  global translations ($\xi^i = a^i= const$), we 
find the conservation law for the canonical stress-energy tensor 
\begin{equation}\label{52} 
\tau^k_{\ i,k} = 0 
\end{equation} 
 with 
\begin{equation} \label{53} 
\tau^k_{\ i} = \frac{\partial \mathcal L}{\partial \phi_{,k}} \phi_{,i} 
- \delta^k_i \mathcal L - (\frac{\partial \mathcal L}{\partial
  \phi_{,k,l}})_{,l}  \phi_{,i} + 
\frac{\partial \mathcal L}{\partial \phi_{,k,l}} \phi_{,i,l}. 
\end{equation}
Invariance under 
 general linear transformations ($\xi^i = \epsilon^i_{\ k} x^k$, with 
constant $\epsilon^i_{\ k}$), determines again the form of the 
stress-energy tensor  
\begin{equation} \label{54} 
\tau^k_{\ i} = 
- \left[\frac{\partial \mathcal L}{\partial \phi_{,k,l}} \phi_{,i}\right]_{,l}
- \frac{1}{2}
\left[
\frac{\partial \mathcal L}{\partial \phi_{,l}} (\sigma \phi)^k_{\ i}
\right]_{,l}
-  \left[\frac{\partial \mathcal L}{\partial \phi_{,m,l}}
[(\sigma \phi)^k_{\ i}]_{,m}\right]_{,l} 
 + \frac{1}{2} \left[\frac{\partial \mathcal L}{\partial \phi_{,m,l}} 
(\sigma \phi)^k_{\ i}\right]_{,m,l}. 
\end{equation}
Invariance under second order transformations ($\xi^i = \epsilon_{ik} x^i
x^k$)  leads to 
\begin{equation} \label{55}
\left[
 \frac{\partial \mathcal L}{\partial \phi_{,k,m}} \phi_{,i} 
+\frac{1}{2} \frac{\partial \mathcal L}{\partial \phi_{,m}} 
(\sigma \phi)^k_{\ i} 
 + \frac{\partial \mathcal L}{\partial \phi_{,l,m}} 
[(\sigma \phi)^k_{\ i}]_{,l} 
\right]_{(km)} = 0 
\end{equation}
where the subscript $(km)$ means that the expression in brackets 
has to be symmetrized in $km$ (no differentiation). 
Finally,  the requirement of general 
covariance ($\xi^i(x)$ arbitrary) leads to 
\begin{equation} \label{56}
\left[
\frac{\partial \mathcal L}{\partial \phi_{,m,l}} (\sigma \phi)^k_{\ i} 
\right]_{(kml)}
= 0, 
\end{equation}
where the expression is totally symmetrized in $kml$ (no differentiation).  
Quite obviously, (\ref{55}) shows that the three first terms 
in (\ref{54}) are 
in the form of a relocalization term $C^{kl}_{\ \  i,l}$, with 
$C^{kl}_{\ \ i} = - C^{lk}_{\ \ i}$, 
while (\ref{56}) shows that the last term has  
the form of a second order relocalization term  
$C^{klm}_{\ \ \ \ i ,l,m}$, with $C^{(klm)}_{\ \ \ \  \ i} = 0$.   
 Thus, $\tau^k_{\ i}$ is identically conserved and $P_k$ can be written 
as a two dimensional surface integral, in accordance with the second 
Noether theorem.  

From (\ref{54}), it is also clear why the Belinfante tensor cannot vanish 
in second order theories. That is because the Belinfante symmetrization 
procedure is based on the definition (\ref{16}). More precisely, the 
Belinfante procedure for second order theories (see \cite{jackiw}) 
consists in writing (\ref{54}) in the form 
\begin{equation} \label{57}
\tau^k_{\ i} = - \frac{1}{2} \tilde L^{mk}_{\ \ i,m},  
\end{equation}
with 
\begin{equation} \label{58}
\tilde L^{mk}_{\ \ i} = 
2 \frac{\partial \mathcal L}{\partial \phi_{,k,m}} \phi_{,i}
+
\frac{\partial \mathcal L}{\partial \phi_{,m}} (\sigma \phi)^k_{\ i}
+ 2  \frac{\partial \mathcal L}{\partial \phi_{,j,m}}
[(\sigma \phi)^k_{\ i}]_{,j}
 - \left[\frac{\partial \mathcal L}{\partial \phi_{,j,m}} 
(\sigma \phi)^k_{\ i}\right]_{,j}, 
\end{equation}
and then defining (compare (\ref{16})) the Belinfante tensor 
\begin{equation} \label{59}
T^{ik} = \tau^{ik} + \frac{1}{2} [\tilde S^{ikm} - \tilde S^{mki} - 
\tilde S^{kmi}]_{,m}, 
\end{equation} 
where $\tilde S^{mki} = \tilde L^{m[ki]}$ and  
$\tilde L^{mki} = \eta^{il} \tilde L^{mk}_{\ \ l}$. This tensor is easily 
shown to be symmetric if the theory is Poincar\'e  invariant, i.e., 
if at least the antisymmetric part of (\ref{54}) is satisfied. Moreover, 
if the theory is invariant under the affine group, i.e., if (\ref{54}) 
is satisfied, then $T^{ik}$ can be written in the form of equation 
(\ref{25}), i.e., 
\begin{eqnarray} \label{60}
T^{ik} &=& - \frac{1}{4}[\tilde L^{mik} - \tilde L^{ikm} + \tilde 
L^{imk} + \tilde L^{mki} + 
\tilde L^{kmi} - \tilde L^{kim}]_{,m} \nonumber \\   
&=& - \frac{1}{2}[\tilde L^{(mi)k} - \tilde L^{(ik)m}  
 + \tilde L^{(mk)i}]_{,m}.     
\end{eqnarray}
If the theory is generally covariant, then we find from (\ref{55}) 
for the symmetric part of $\tilde L^{mki} $ the following relation 
\begin{equation} \label{61}
\tilde L^{(mk)i} = 
- \frac{1}{2}\left[\frac{\partial \mathcal L}{\partial \phi_{,j,m}} 
(\sigma \phi)^{ki}\right]_{,j} 
- \frac{1}{2}\left[\frac{\partial \mathcal L}{\partial \phi_{,j,k}} 
(\sigma \phi)^{mi}\right]_{,j}, 
\end{equation} 
from which you can explicitely evaluate $T^{ik}$. It does not 
identically vanish, but it is nevertheless identically conserved, 
as a result of  (\ref{56}). 
In the case of general relativity, 
with the covariant, second order Lagrangian 
$\mathcal L = \frac{1}{2}\sqrt{-g}R$,  
it leads again to the expression (\ref{43}), as is easily verified, 
in accordance with the results of  \cite{jackiw}\footnote{It 
is understood that the matter part of the Lagrangian is first order, 
and thus does not explicitely contribute to $T^{ik}$. If this is not 
the case, you will have, apart from (\ref{43}),  
 additional terms in $T^{ik}$, which, however, 
do not contribute to $P_k$ if the usual asymptotical behavior is 
assumed. A similar assumption has been made in \cite{jackiw}, where 
the Belinfante procedure was applied to the canonical tensor of the 
gravitational field, and then brought into the form (\ref{43}) by the 
use of the gravitational field equations. The result is then the 
Belinfante tensor of gravity plus the Hilbert tensor of matter. 
However, the Hilbert tensor is not equal to the Belinfante tensor 
of the matter field if the matter Lagrangian contains  second order
derivatives (see next section). The only generally valid form for the total 
Belinfante tensor  $T^{ik}$ is therefore 
(\ref{60}) with (\ref{61}).}. It is interesting to remark that the 
evaluation of $T^{ik}$ involves a lot less computation than that of 
$\tau^i_{\ k}$. 

We wish to point out, however, that you can always write the last 
term in (\ref{54}) in the form of a first order relocalization term too.
The has been shown in \cite{goldberg} for the general case. 
Specifically, in general relativity, this term is equal to 
\begin{displaymath}
- \frac{1}{2}\left[\sqrt{-g}\ (-2 g^{lm} \delta^k_i + g^{lk} \delta^m_i + 
g^{km} \delta^l_i)\right]_{,m,l},  
\end{displaymath} 
which corresponds to the second line in (\ref{42}) (hence 
the identical Belinfante tensors in first and second order theory). 
An equivalent form of this expression is 
\begin{displaymath}
\left[\sqrt{-g}\ ( g^{lm} \delta^k_i - 
g^{km} \delta^l_i)\right]_{,l,m},  
\end{displaymath} 
which is now of the form $-\frac{1}{2} \hat L^{lk}_{\ \ i,l}$, 
with $\hat L^{lk}_{\ \ i}$ antisymmetric in $kl$. Together with the 
three first terms in (\ref{54}), we can now write  
the stress-energy in the form $\tau^k_{\ i} = 
- \frac{1}{2}  \overline L^{lk}_{\ \ i,l}$, with antisymmetric 
$ \overline L^{lk}_{\ \ i}$, i.e., in the form of a first order 
relocalization term (although still equal to (\ref{57})). There is no    
reason to prefer one or the other form, meaning  that 
the generalization of the Belinfante procedure to second order 
theory is not really as unambiguous as it has been presented in \cite{jackiw}. 
Indeed, defining the Belinfante tensor using $\overline L^{lki}$ in 
(\ref{60}) leads to a vanishing tensor.  

Finally, we note that (\ref{61}) reduces in the case of 
first order theories to (\ref{29}), which completes the  proof 
of the results of section 4 without explicit reference to the vector or 
tensor nature of the fields.   

The above results do 
not mean that there is any fundamental difference between 
first and second order theories. In both kind of theories, the 
stress-energy tensor is given in the form of relocalization terms 
and can thus obviously be annihilated  by adding appropriate 
relocalization terms, of first or second order. Also, in both cases, 
relocalization terms modify the momentum vector in the case of 
the linear Hilbert-Einstein or Einstein-Cartan type theories. 
The only difference 
comes from the following: In first order theories, if you eliminate 
the antisymmetric part of $\tau^{ik}$, making it thus suitable for 
the formulation of conservation laws for the full Poincar\'e  
group, then you automatically annihilate the complete tensor, while 
in second order theories, you can eliminate  the antisymmetric part and 
still have a non-vanishing tensor. However, as we have argued before, 
in generally covariant theories, it is rather arbitrary to pick out 
the Poincar\'e group and modify the stress-energy tensor accordingly. 
There seem to be only two reasonable (albeit extreme) 
points of view: Either, we allow 
for modifications of $\tau^i_{\ k}$ by 
relocalization terms or we don't. If we do, then we should use them  
in order to construct the tensor that has the largest possible symmetry. 
This is quite obviously the tensor that vanishes identically, since it is 
trivially fully covariant. If we do not, 
then we should simply stick to the canonical tensor in its initial form. 
Everything else is, in the framework of generally covariant 
theories, an ad hoc procedure and can only be of limited usefulness. 
(See, however, the remarks at the end of section 5.)

\section{Matter stress-energy and the Hilbert tensor}  

In this section, we split the Lagrangian into a gravitational and a 
matter part, $\mathcal L = \mathcal L_0 + \mathcal L_m$, and briefly examine 
the relations between the canonical stress-energy for the matter fields, 
the corresponding Belinfante tensor and the so-called Hilbert tensor. 
Since the 
relation between canonical and Hilbert tensor has been discussed  in  
detail in our previous article \cite{leclerc}, and since the 
generalization of the discussion to the Belinfante tensor is rather 
simple, in view of the relations we have already obtained in the previous 
sections, we will confine ourselves to give a very short exposition of the 
issue. The subject has also been covered, e.g.,  in \cite{borokhov} and 
\cite{saravi}. 

We consider purely metric theories (e.g., general relativity), 
and for (some kind of) 
completeness, we allow, for the moment, derivative couplings 
(first order) of the metric to the matter fields. Nevertheless,  non-minimal 
couplings of the form $\sqrt{-g}\ \phi^2 R$, are still excluded, since they 
contain second derivatives of the metric coupling to $\phi$.  
Moreover, we confine ourselves to 
first order theories, i.e., we assume that 
no second derivatives of the matter fields 
occur in the Lagrangian. 

The variation of the matter Lagrangian then is of the form 
\begin{equation} \label{62} 
\delta \mathcal L_m = 
\frac{\partial \mathcal L_m}{\partial \phi} \delta \phi 
+ \frac{\partial \mathcal L_m}{\partial \phi_{,i}} \delta \phi_{,i} 
+ \frac{\partial \mathcal L_m}{\partial g_{lm}} \delta g_{lm} 
+ \frac{\partial \mathcal L_m}{\partial g_{lm,i}} \delta g_{lm,i},  
\end{equation}
where $\phi$ denotes collectively the matter fields (i.e., all the 
dynamical fields, except for $g_{lm}$). Note that the matter fields 
are supposed to satisfy the field equations, 
while for $g_{lm}$, no field 
equation can be used, since the gravitational part $\mathcal L_0$ is 
not included  in our Lagrangian. In other words, $g_{lm}$ is treated as 
background field. Under a  coordinate transformation, the 
fields transform according to  (\ref{51}), and 
in particular,  for the metric we 
have $\delta g_{lm} = g_{lm,k} \xi^k + \xi^k_{\ ,l} g_{km} 
+ \xi^k_{\ ,m} g_{lk}$. The matter Lagrangian is assumed to be a 
scalar density, i.e., $\delta \mathcal L_m = (\mathcal L_m \xi^k)_{,k}$. 

The Hilbert stress-energy tensor is defined by 
\begin{equation}\label{63}
\mathcal T^{lm} = - 2  \frac{\delta \mathcal L_m}{\delta g_{lm}} 
= -2 
\frac{\partial \mathcal L_m}{\partial g_{lm}}
+2 
\left(\frac{\partial \mathcal L_m}{\partial g_{lm,i}}\right)_{,i},  
\end{equation}
where, in accordance with our previous conventions, we have defined 
$\mathcal T^{lm} $ as tensor density. In view of the invariance of 
$\mathcal L_m$ under coordinate transformations, $\mathcal T^{lm}$ satisfies 
the covariant conservation law $\mathcal T^{lm}_{\ \ ;l}= 
\mathcal T^{lm}_{\ \ ,l} + \Gamma^m_{kl} \mathcal T^{lk} = 0$. 
The derivation of this law is found in any textbook on general 
relativity (or, see  \cite{leclerc}). Note however that it holds independently 
of the specific form of the gravitational Lagrangian. 

With this definition, and using $\mathcal T^{lm}_{\ \ ;l} = 0$, 
we derive for $\xi^i = a^i$ the 
following conservation law 
\begin{equation} \label{64}
\tau^i_{\ k,i} - \mathcal T^i_{\ k,i} + 
\left[\frac{\partial \mathcal L_m}{\partial g_{lm,i}}\right]_{,i} = 0,  
\end{equation}
where $\tau^i_{\ k}$ is defined as in (\ref{9}) (but with $\mathcal L_m$), 
and $\mathcal T^i_{\ k} = \mathcal T^{im}g_{mk}$. 
Next, for linear transformations, 
we get 
\begin{equation} \label{65} 
\tau^i_{\ k} - \mathcal T^i_{\ k} + \frac{1}{2}  L^{mi}_{\ \ k,m} 
+ 2 \left[ \frac{\partial \mathcal L_m}{\partial g_{li,m}} g_{kl}
\right]_{,m} +  \frac{\partial \mathcal L_m}{\partial g_{lm,i}} g_{lm,k} = 0, 
\end{equation} 
where $ L^{mi}_{\ \ k} $ is defined as in (\ref{22}). 
Requiring general covariance, we find the additional relation 
\begin{equation} \label{66}
 \frac{1}{2}  
L^{(mi)}_{\ \ \ k} 
+ 2  \frac{\partial \mathcal L_m}{\partial g_{l(i,m)}} g_{kl} = 0, 
\end{equation}
showing that 
$\tau^i_{\ k} - \mathcal T^i_{\ k} +
\frac{\partial \mathcal L_m}{\partial g_{lm,i}} g_{lm,k} $ 
 is identically conserved. It is not hard to see from the  definition 
(\ref{16}) that the Belinfante tensor is not equal to $\mathcal T^{ik}$, 
nor does it reduce to $\mathcal T^{ik}$ in the flat limit $g_{ik} =
\eta_{ik}$. However, in the case where derivative couplings are absent, 
(\ref{65}) simplifies to 
$\tau^i_{\ k} - \mathcal T^i_{\ k} = - \frac{1}{2}  L^{mi}_{\ \ k,m}$, 
and, since (\ref{66}) reduces  to $L^{(mi)}_{\ \ \ k} = 0$, 
 the Belinfante tensor $T^{ik}$ is now easily shown to be directly related 
to $\mathcal T^{ik}$, namely, we have 
\begin{equation} 
T^{ik} = \mathcal T^i_{\ m} \eta^{mk}, 
\end{equation}  
where we recall that $\mathcal T^i_{\ m} = g_{mk} \mathcal 
T^{ik}$, and thus, since the 
second index of $T^{ik}$ is lowered with the Minkowski metric, we can 
equivalently write 
\begin{equation} 
T^i_{\ k}  = \mathcal T^i_{\ k}. 
\end{equation}
In summary, in first order theories without derivative couplings, 
the Belinfante tensor is indeed equal to the Hilbert tensor (when 
written in mixed form). Note that the Belinfante (matter) tensor $T^{ik}$ is, 
strictly speaking, not symmetric (since it is not equal to $\mathcal T^{ik}$ 
in contravariant form). A symmetric tensor is obtained by 
raising the second index of $T^i_{\ k}$ with $g^{ik}$ instead of $\eta^{ik}$. 
This, however, is not really of importance, since $T^i_{\ k}$ is not 
conserved anyway in a curved background (neither is $\mathcal T^{ik}$), 
and in general, neither will the angular momentum be conserved. 

The equality of the Belinfante and the Hilbert tensors in first order 
theories in the framework of metrical theories of gravity 
can give rise to two different interpretations. The first one 
is  that it provides a strong argument to the standpoint 
that  
the Belinfante procedure can indeed be used  to 
derive (or rather define)  the {\it correct} stress-energy tensor not 
only in special relativity, but also in metric theories of gravity, 
and in particular for the gravitational Lagrangian itself. A second, 
quite different point of view  consists in interpreting the Hilbert 
tensor as a covariant generalization of the Belinfante tensor, a point of 
view equally strongly supported by the above equality. According to the 
second interpretation, the stress-energy tensor of the gravitational field 
itself is then given by the variation of $\mathcal L_0$ with respect to the 
metric, and quite obviously, the total stress-energy tensor would 
vanish in view of the gravitational field equations. 
This interpretation is supported by 
the fact that in generally covariant first order theories, the total 
Belinfante tensor is indeed identically zero, while in second order 
theories (like general relativity), one could modify the Belinfante 
symmetrization procedure, including second order relocalization terms, 
in a way that again, the total tensor would vanish. 

We have carried out a similar analysis of the relation between the 
canonical and the corresponding (generalized Hilbert) 
tensor $\mathcal T^i_{\ a} 
= - \frac{\delta \mathcal L_m}{\delta e^a_i}$ in 
the framework of Einstein-Cartan theory and also in tetrad gravity 
in \cite{leclerc}. It was shown in particular that both tensors 
are not in general related by a relocalization term (because of the additional 
field $\Gamma^{ab}_{\ \ i}$ in the first case, and because of 
the derivative couplings to spinor fields in the second case), and it is 
not hard to extend the analysis to include the Belinfante tensor. The result 
is that there is no equality (not even up to a relocalization term) 
in those theories between the Hilbert and the Belinfante tensors.   
For instance, in Einstein-Cartan theory, one can derive 
the relation $T^i_{\ k} = \mathcal T^i_{\ a} e^a_k - \sigma_{ab}^{\ \ i}
\Gamma^{ab}_{\ \ k}$, with $\sigma_{ab}^{\ \  i} = 
\partial \mathcal L_m/\partial \Gamma^{ab}_{\ \ i}$.  
Equality holds in the absence of spinor fields 
$\sigma_{ab}^{\ \ i} = 0$ or in the limit of vanishing gravity,    
$e^a_i= \delta^a_i, \Gamma^{ab}_{\ \ k} = 0$. Similarly, in tetrad 
gravity, we have equality only in those limits. 

Moreover, the above result for metric theories does not generalize to 
matter Lagrangians containing second derivatives. This is actually quite 
obvious, since if it would, then the same would be true for the gravitational 
part of the Lagrangian, and then, the total Belinfante tensor 
(adopting the definition of \cite{jackiw}, i.e., using 
(\ref{58}) and (\ref{60})) 
would 
be equal to the variation of the total Lagrangian with respect to $g_{ik}$, 
and thus zero, which is, however, not the case in second order theories, 
as we have shown in the previous section. 
The  equality between Hilbert and 
Belinfante tensor can  be achieved by 
modifying the Belinfante relation for second order theories, 
as we have done in the previous section, 
rewriting the  second order relocalization 
term in the form of a first order term, such that the total Belinfante 
tensor vanishes again. 
However, in view of the physical irrelevance of matter Lagrangians with 
second derivatives, we do not pursue this idea further.

\section{Generators and Hamiltonian constraints} 

In this section, we briefly review the relation between the conserved 
charges, the generators of the corresponding symmetries and the 
(first class) Hamiltonian constraints. The issue is of fundamental 
importance for the quantization of field theories with local symmetries 
and has been pioneered by Bergmann and Dirac. We confine ourselves to 
a brief analysis of the important cases (internal gauge symmetry 
 and general covariance) and refer the reader to the initial articles, 
in particular \cite{bergmann1}, \cite{bergmann2}, \cite{bergmann3}, 
\cite{fletcher} and \cite{dirac}. 

\subsection{Gauge theory}

In order to present our equations in a covariant form, we use the 
covariant formalism of \cite{leclerc2}. In particular, we will 
use integrals over spacelike hypersurfaces $\sigma$ defined by 
$\Phi(x) = 0$, with (timelike) normal vector $n_i = \Phi_{,i}$. The normal 
vector is understood to be  normalized,  $g^{ik} n_i n_k = 1$. Note 
that this normalization is only a convention used for greater convenience 
in intermediate steps. Our initial and final relations should not depend 
on this normalization, since we do not want to assert any special meaning 
to the metric $g_{ik}$, which is considered to be a field like any other
field. 

Further, we define the canonical momenta by 
\begin{equation} \label{69}  
\pi^{(i)} = \frac{\partial \mathcal L}{\partial \phi_{,i}}, 
\end{equation}
where we put the index into parentheses to remind of the fact that only 
the part normal to the hypersurface, $\pi = \pi^{(i)} n_i$ corresponds 
to the physical momentum known from conventional Hamiltonian theory. 

We start again with non-abelian gauge theory. The conserved current 
density has been derived in the form 
\begin{equation} \label{70} 
J^{  i}_{\alpha} =  
-i \frac{\partial \mathcal L}{\partial \psi_{,i}} \sigma_{\alpha} \psi 
+ \frac{\partial \mathcal L}{\partial A^{\beta}_{k,i}} 
c_{\alpha\gamma}^{\ \  \beta} A^{\gamma}_k
=
-i \pi^{(i)}_{\psi} \sigma_{\alpha} \psi 
+ \pi^{k(i)}_{\beta} 
c_{\alpha\gamma}^{\ \  \beta} A^{\gamma}_k
\end{equation}
Next, in view of equation (\ref{3}) we define\footnote{Note that 
this is the opposite of the expression (\ref{3}), if (\ref{4}) is 
taken into account.} 
\begin{equation} \label{71} 
\tilde J^i_{\alpha} = - 
(\frac{\partial \mathcal L}{\partial A^{\alpha}_{k,i}})_{,k} 
=- \pi^{k(i)}_{\alpha\ \ ,k} 
\end{equation}
and in view of equation (\ref{4}), 
\begin{equation} \label{72}
\hat J^{ki}_\alpha = \frac{\partial \mathcal L}{\partial 
A^{\alpha}_{i,k}}  = \pi^{i(k)}_{\alpha},  
\end{equation}
as well as the corresponding charges 
\begin{equation} \label{73}
Q = \int \epsilon^{\alpha} J^i_{\alpha} \de \sigma_i, \ \ \ 
\tilde Q = \int \epsilon^{\alpha} \tilde J^i_{\alpha} \de \sigma_i, \ \ \  
\hat Q = \int [\epsilon^{\alpha}J^{ki}_{\alpha} ]_{,k} \de \sigma_i,  
\end{equation}
where $\epsilon^{\alpha}(x)$ is an arbitrary parameter. 

As a result of equation (\ref{3}), we have the identically satisfied 
relation $J^i_{\alpha} + \tilde 
J^i_{\alpha} = 0$, and as a result of (\ref{4}), 
$\hat J^{ik}_{\alpha} + \hat J_{\alpha}^{ki}= 0$. 
It is not hard to 
recognize that the corresponding equation 
$Q + \tilde Q = 0$ is equivalent to the secondary (first class) constraints 
of the theory. Consider, e.g., flat space and choose $n_i = \delta^0_i$ (that
is, the hypersurface $t = const$). 
Then, the integrand of $Q+ \tilde Q$ reduces  to 
$\epsilon^{\alpha}[- i \pi_{\psi} \sigma_{\alpha} \psi + \pi_{\beta}^k 
c_{\alpha\gamma}^{\ \ \beta} A_k^{\gamma} - \pi_{\alpha\ ,\mu}^{\mu}]$, 
where $\mu = 1,2,3$. In the last term, we have exploited the antisymmetry 
of $\pi_{\alpha}^{k(i)}$ resulting from (\ref{4}). This is exactly the 
secondary constraint arising in Yang-Mills-Dirac theory. 
In the context of quantization in the Coulomb gauge, it is usually referred to 
as Gauss' law. More generally, the relation 
$J^i_{\alpha} + \tilde J^i_{\alpha}= 0$ is equivalent to the field 
equations for the Yang-Mills field, but the fact that  it can be derived 
directly from the symmetry of the theory indicates the presence of 
constraints. The field equations for $\psi$ for instance cannot be derived in 
this way. 

In a similar way, the relation 
$\hat J^{(ik)}_{\ \alpha} = 0$ is related to the primary constraints. 
Consider again the case $n_i = \delta_i^0$. Then, we have 
$\pi^{k}_{\alpha} =  \pi^{k(0)}_{\alpha}$, and therefore $\pi^{0}_{\alpha} 
= 0$, which is the well know primary constraint of Yang-Mills theory.  
Alternatively, with the help of equation (\ref{app4}) of Appendix A, we can 
express the primary constraints simply as $\hat P = 0$.

Next, we analyze the relations of $Q$, $\tilde Q$ and $\hat Q$ to 
the generators of the gauge transformation. For this, we assume 
the following commutation relations between fields and canonical momenta 
\begin{equation} \label{74} 
[ \phi(x), \phi(y)] = 0, \ \ [\pi^{(i)}(x),\pi^{(m)}(y)] = 0, \ \ 
[\phi(x), \pi^{(i)}(y)]= \delta_{\sigma}^i(x-y), 
\end{equation}
where $x,y$ are assumed to be separated by a 
spacelike distance. For details on  
our formalism, we refer to  Appendix A. 
In the case of spinor fields, the above relations are replaced by 
anticommutation relations, e.g., $\{\psi^{M}(x), \pi^{(i)}_N(y)\} 
= \delta^M_N \delta^i(x-y)$, where $M,N$ denote the  spinor components. 
It is not of our concern here where the (anti)commutation relations 
come from, be it from a classical Poisson bracket (see \cite{leclerc2}) or 
be it from postulating them in a quantum theory. They are simply assumed 
to be valid. 

We are now ready to evaluate the commutation relations between the 
fields and the charges defined in (\ref{73}). For the manipulations 
involved,   a few helpful rules are provided in Appendix A.   
Here, we give only the results. Note that in order to derive those 
results, two dimensional surface terms have been omitted during the 
calculations. This is permitted if we assume that the parameter 
$\epsilon^{\alpha}$ has an appropriate asymptotical behavior, or 
if we simply assume that it vanishes outside of a certain region. 

We find the following relations 
\begin{eqnarray}\label{75}
[Q, A_l^{\beta}] &=& - \epsilon^{\alpha} c_{\alpha\gamma}^{\ \ \beta} 
A^{\gamma}_l, \ \ \ 
\ \ \ \ \ \ \ \ \ \ \ \ \ \ 
\ \ \ \ \ \  \ \ \  \ \ \ \ \ \ \  \ \ \ \ \ \ 
[Q, \psi] =  i \epsilon^{\alpha} \sigma_{\alpha} \psi, \\ \label{76} 
[\tilde Q, A_l^{\beta}]&=&  -\epsilon^{\beta}_{,l} + \epsilon^{\beta}_{,m} 
n^m n_l + \int \epsilon^{\beta} [\delta^i_{\sigma}(x-y)]_{,i} \de \sigma_l, 
\ \ \ [\tilde Q, \psi] = 0, \\ \label{77} 
[\hat Q,A^{\beta}_l] &=&
 - \epsilon^{\beta}_{,m} n^m n_l
- \int \epsilon^{\beta} [\delta^i_{\sigma}(x-y)]_{,i} \de \sigma_l, 
\ \ \ \ \ \ \ \ \ \  [\hat Q, \psi] =0.
\end{eqnarray}
In order not to overload the notation, we have omitted the arguments 
of the fields. It is understood that all the fields 
(including $\epsilon^{\alpha}$) are taken at 
a point $y$, lying on the hypersuface with respect to which the 
charges have been defined in (\ref{73}), except in the 
second terms of (\ref{76}) and (\ref{77}), where $\epsilon^{\alpha}$ 
is taken at the point $x$, and the integration is performed over 
$\de \sigma_l(x)$. As pointed out in Appendix A, there is no 
obvious way to simplify those integrals without  specifying a hypersurface, 
but they do not appear in 
the combinations of the generators we are interested in. 
If we choose the hypersurface $t = const$ 
for $\mu = 1,2,3$, we find the more conventional 
form\footnote{Note that the same results can be directly derived by the use 
of the more familiar, but not covariant generators 
\begin{displaymath}
Q = \int \epsilon^{\alpha} J^0_{\alpha} \de^3 x, \ \ \ 
\tilde Q = \int \epsilon^{\alpha} \tilde J^0_{\alpha} \de^3 x \ \ \ 
\hat Q = \int [\epsilon^{\alpha} \hat J^{00}_{\alpha}]_{,0} \de^3 x, 
\end{displaymath} 
instead of (\ref{73}),  
where we identify directly the physical momenta with $\pi= \pi^{(0)}$, 
and take  the commutation relations to be 
  (at equal times)
 $[\phi(x), \pi(y) ] = \delta^{(3)}(\vec x-\vec y) $.}  
\begin{displaymath}
[\tilde Q, A^{\beta}_{\mu}] = - \epsilon^{\beta}_{,\mu}, \ \ \  
[\hat Q, A^{\beta}_{0}] = - \epsilon^{\beta}_{,0}, 
\end{displaymath} 
as well as  $[\tilde Q,A^{\beta}_{0} ]= [\hat Q, A_{\mu}^{\beta}] = 0.$  

Those results are very interesting. First, we see that the charge 
$Q$ does not (as is occasionally  stated)  generate gauge transformations 
on the fields. Rather, it generates homogeneous   transformations on 
both $A_l^{\beta}$ and $\psi$, in the  adjoint (or vector) 
representation  on $A_l^{\beta}$ (a rotation in isospin space) and 
in the fundamental representation  on $\psi$. 

The inhomogeneous part is generated by $\tilde Q + \hat Q$, namely we have  
\begin{equation} \label{78}
[ \tilde Q + \hat Q, A^{\beta}_l] = -\epsilon^{\beta}_{,l}. 
\end{equation}
In order to find the generator of the full  gauge transformation, we 
need all three parts, i.e., 
\begin{equation} \label{79}
[ Q +\tilde Q + \hat Q, A^{\beta}_l] = -\epsilon^{\beta}_{,l} 
- \epsilon^{\beta}_{\ \gamma \alpha} A^{\gamma}_l, \ \ \ 
[Q+ \tilde Q + \hat Q, \psi] =  i \epsilon^{\alpha} \sigma_{\alpha} \psi,
\end{equation} 
which is exactly the transformation (\ref{1}) we started from. 
Both (\ref{78}) and (\ref{79}) are independent of the hypersurface. 
Another 
combination of interest is the generator $Q + \tilde Q$, because 
it is identically zero (related to the secondary constraint). 
If we choose the hypersurface $t = const$, we find for the 
spatial components $[Q + \tilde Q, A^{\beta}_{\mu}] = -\epsilon^{\beta}_{,\mu}
- \epsilon^{\beta}_{\ \ \gamma\alpha} A^{\gamma}_{\mu}$. The fact that 
an  operator that vanishes weakly (i.e., as a result of the constraints)
generates   gauge transformations 
on the propagating fields (consider, e.g., the Coulomb gauge), is interpreted 
in quantum theory as the expression for the fact that physical states have 
to be singlets under the gauge group. That is, free bosons $A_{\mu}^{\beta}$ 
cannot be  part of the physical particle spectrum (see, e.g., \cite{kaku}). 
This should be somehow alerting, considering the fact that we will now 
go over to spacetime symmetries.

\subsection{General covariance} 

The analysis is quite similar to the previous case. In view of the 
relations (\ref{9}), (\ref{23}) and (\ref{29}), we define\footnote{Note 
that again, the second expression (\ref{81}) corresponds to the opposite 
of (\ref{23}).} 
\begin{equation} \label{80} 
\tau^i_{\ k} = \pi^{(i)} \phi_{,k} - \delta^i_k \mathcal L, 
\end{equation} 
 \begin{equation} \label{81} 
\tilde \tau^i_{\ k} = -\frac{1}{2} L^{im}_{\ \ k,m}=  
 - \frac{1}{2} [\pi^{(i)} (\sigma \phi)^m_{\ k}]_{,m}, 
\end{equation}
and 
\begin{equation} \label{82} 
\hat \tau^{mi}_{\ \ k} =  \frac{1}{2} 
L^{mi}_{\ \ k} = \frac{1}{2} \pi^{(m)} (\sigma \phi)^i_{\ k}. 
\end{equation}
Next, we introduce the corresponding charges\footnote{Again, there is 
the more familiar choice $P = \int \epsilon^k \tau^0_{\ k} \de^3 x$, 
$\tilde P  = \int \epsilon^k \tilde \tau^0_{\ k} \de^3 x$ and 
$ \hat P = \int [\epsilon^k \hat \tau^{00}_{\ \ k}]_{,0} \de^3 x$, 
involving only the components  $\pi^{(0)}$ of the momenta, which  
 satisfy $[\phi(x), \pi(y)] 
= \delta^{3}(\vec x - \vec y)$.}

\begin{equation} \label{83} 
P = \int \epsilon^k \tau^i_{\ k} \de \sigma_i, \ \ \ 
\tilde P = \int \epsilon^k \tilde \tau^i_{\ k} \de \sigma_i, \ \ \ 
\hat P = \int [\epsilon^k \hat \tau^{mi}_{\ \ k}]_{,m} \de \sigma_i.  
\end{equation}
The primary constraints $\hat P = 0$ arise again from the fact that 
$\hat \tau^{mi}_{\ \ k}$ 
is antisymmetric in $mi$,  
while the secondary constraints are expressed 
by $P + \tilde P = 0$. The explicit form will depend on the nature of 
the field, i.e., on the form of $(\sigma \phi)^i_{\ k}$. Note also 
that the relation $\tau^i_{\ k} + \tilde \tau^i_{\ k} = 0$ is 
equivalent to a field equation. In general relativity, this would 
be the Einstein equation. In other theories, it could be a combination
of field equations.

In order to evaluate the commutation relations with $\phi$, it is 
important to recall that in (\ref{80}), the Lagrangian is expressed in 
terms of the field and its derivatives. It is not understood that 
field derivatives are to be replaced by momenta. This cannot be done 
in a unique way anyway, since (\ref{80}) is not a Legendre transformation. 
In other words, $\tau^i_{\ k}$ has nothing to do with the field 
Hamiltonian, which is indeed a Legendre transformation 
(see \cite{leclerc2} for more details and on the definition of the 
Hamiltonian in the manifestly covariant formalism).  

We find the following commutation relations 
\begin{eqnarray}\label{84}
[P, \phi] &=& -\phi_{,k} \epsilon^k, \\  \label{85}
[\tilde P, \phi] &=& -\frac{1}{2} \epsilon^k_{,m} (\sigma \phi)^m_{\ k} 
 + \frac{1}{2} [(\sigma \phi)^m_{\ k}\epsilon^k]_{,i} n_m n^i
+ \frac{1}{2} \int [\delta^i_{\sigma}(x-y]_{,i} 
(\sigma \phi)^m_{\ k} \epsilon^k 
\de \sigma_m, \\  \label{85b}
[\hat P, \phi] &=& -\frac{1}{2} [ (\sigma \phi)^k_{\ m}\epsilon^m]_{,i} n_k n^i
- \frac{1}{2} \int [\delta^i_{\sigma}(x-y)]_{,i} 
(\sigma \phi)^m_{\ k} \epsilon^k 
\de \sigma_m. 
\end{eqnarray}
In particular, for the sum $\tilde P + \hat P$, we have 
\begin{equation}\label{86}
[\tilde P + \hat P, \phi] = -\frac{1}{2} \epsilon^k_{,m} 
(\sigma \phi)^m_{\ k}. 
 \end{equation} 
 The situation is in complete analogy to the previous case. The 
charge $P$ (i.e., the canonical field momentum in the strict sense)   
generates what is usually referred to as spacetime translations, in the 
sense that it tells us the evolution of $\phi$ from one point to another, 
$\phi(x) - \phi(x') = -\phi_{,k} \epsilon^k$. On the other hand, the 
operator $\tilde P + \hat P$ generates passive coordinate transformations 
as they are usually considered in general relativity, i.e., 
$\phi(x)- \phi'(x') = -\frac{1}{2} \epsilon^k_{,m} (\sigma \phi)^m_{\ k}$. 
And finally, the operator $P + \tilde P + \hat P$ generates 
the active coordinate transformations we started from, i.e., the 
Lie derivatives of the field 
\begin{equation}\label{87}
[P + \tilde P+ \hat P, \phi] 
= -\phi_{,k} \epsilon^k - \frac{1}{2} \epsilon^k_{,m} (\sigma \phi)^m_{\ k} 
 = \phi'(x) - \phi(x), 
\end{equation}
with $x'^i = x^i+ \epsilon^i$.  

A few remarks are in order at this point. The relation (\ref{84}) is 
in the form one expects for a momentum operator. It is the operator 
one wants to have in a quantum theory. And it is based on the 
canonical stress-energy tensor. Now, since 
in the framework of general relativity, $\tau^i_{\ k} + \tilde \tau^i_{\ k} 
= 0$ is equivalent to Einstein's equation, i.e.,  to 
$\delta \mathcal L/ \delta g_{ik} = 0$, the tensor  corresponding 
to the gauge generator $P + \tilde P$ is obviously given by 
$- \sqrt{-g}G^{ik} + \mathcal T^{ik}$, 
where $G^{ik} $ is the Einstein tensor and 
$\mathcal T^{ik}$ the Hilbert tensor (density) 
for the matter fields. In particular, 
concentrating on the matter part, we see that $\mathcal T^{ik}$, which is 
also equal to the Belinfante (matter) tensor $T^{ik}$, 
is related to a gauge transformation of the form 
(for the hypersurface $t = const$) $\delta \phi = -\epsilon^k \phi_{,k} 
- \frac{1}{2} \epsilon^k_{, m} 
 (\sigma\phi)^{m}_{\ k} 
+ \frac{1}{2} [(\sigma \phi)^0_{\ k}\epsilon^k]_{,0}$. 
 This makes it rather hard to interpret  the 
quantity  $\int \mathcal T^0_{\ k} \de^3 x$ as field momentum. Quite 
obviously, this interpretation should be reserved to the corresponding 
expression with the canonical tensor. 

In short, the Belinfante is  not 
directly related to the generators of coordinate transformations  
and as such, its  integral over space should not be directly interpreted as 
energy and momentum. 

A second remark concerns the same generator, $P + \tilde P$, 
which is zero (secondary constraint), 
and generates again gauge transformations 
on the propagating fields, quite similar as in the case of internal  gauge 
theories. For instance,  for the spatial components of a tensor (e.g., the 
metric)   
 we find (for the hypersurface $t = const$) 
$[P + \tilde P, g_{\mu\nu}] = - g_{\mu\nu,k}\epsilon^k 
- \epsilon^k_{\ ,\mu} g_{k \nu} - \epsilon^k_{,\nu}g_{\mu k}$. For 
transformations restricted to the hypersurface, we have 
$\epsilon^k = (0, \epsilon^{\mu})$, and thus, we find that 
$[P+ \tilde P, g_{\mu\nu}]$ is equal to the   
 Lie derivative of $g_{\mu\nu}$ in the three dimensional subspace. 
One is thus tempted to conclude that physical states have to be  
singlets under  coordinate transformations, i.e., scalars. This would 
not only exclude gravitons, but also photons and vector fields in general. 
While for the latter, we can argue that they are considered usually on 
a given background (with a reduced symmetry), in the case of the gravitational 
theory, the solution to this problem is more profound and can be 
found in the fact that general 
relativity (or similar theories, e.g.,  Einstein-Cartan) is actually a 
theory with a spontaneous symmetry breaking, in the sense that the vacuum 
is given by $g_{ik} = \eta_{ik}$ (or eventually a de Sitter metric)
and thus allows only for Poincar\'e (or de Sitter) transformations. This 
invalidates the previous arguments, and gravitons can now be interpreted 
as fields propagating on this background. (Note that similarly, 
massive gauge bosons in the Salam-Weinberg model are also part of the 
physical spectrum.) 

Finally, we caution that we have assumed in our analysis that the theory is 
free of second class constraints. Second class constraints 
arise in particular in 
the Dirac theory, but also in Einstein-Cartan theory (see \cite{blago}) 
and are not related to gauge symmetries. Their presence 
leads to modifications of the canonical (anti)commutation relations 
and  to modifications of the above results, in particular of  
the relation (\ref{84}).  We refer to \cite{leclerc2} for 
details on the Dirac theory. As to Einstein-Cartan theory, one has to 
take into account both the diffeomorphism covariance and the local Lorentz 
invariance (\ref{33}). This leads to the primary constraints 
(identifying directly $\pi$ with $\pi^{(0)}$ for simplicity) 
$\pi_{ab}^0 = 
\pi_a^0 = 0$, where $\pi_{ab}^i$ and $\pi_a^i$ are the momenta canonically 
conjugated to $\Gamma^{ab}_{\ \ i}$ and $e^a_i$, respectively. The remaining 
primary constraints, namely $\pi_a^{\mu} = 0$ and $\pi_{ab}^{\mu} 
+\frac{e}{2}(e^0_a e^{\mu}_b - e^{\mu}_a e^0_b)= 0 $,  
are not related to any kind of symmetry and are second class, i.e., 
the commutators $\left[
\pi^{\nu}_c, \ \pi_{ab}^{\mu}+\frac{e}{2}(e^0_a e^{\mu}_b - 
e^{\mu}_a e^0_b)\right]$ do not all vanish, as can be seen by expressing 
$e e^{\mu}_a$ in terms of $e^a_i$.

\subsection{General relativity}

The previous analysis is valid for  generally covariant  first order 
theories, and as such, is directly applicable to, e.g., Einstein-Cartan 
theory or Poincar\'e gauge theories in general, once we have dealt 
consistently with the second class constraints.  
On the other hand, 
general relativity is either second order, or the Lagrangian is not 
generally covariant  and requires therefore special care. We will not 
deal with the second order theory here, but instead, try to {\it fix} 
the first order theory in order to make it suitable for the above 
analysis. 

For simplicity, we consider pure gravity, such that $g_{ik}$ is the only 
field, and we have 
\begin{equation} \label{88}
L^{mk}_{\ \ i} = \frac{\partial \mathcal L}{\partial g_{ln}} 
(\sigma g_{ln})^k_{\ i} 
= 4 \frac{\partial \mathcal L}{\partial g_{kl,m}} g_{li}
= 4 \pi^{kl(m)} g_{li} . 
\end{equation}
The explicit expression is given in (\ref{42}). In particular, 
as a result of the non-covariance of the Lagrangian (\ref{41}), $L^{mk}_{\ \
  i}$ is not antisymmetric in $mk$. The relation  (\ref{88}) is actually 
quite interesting, since it tells us that there are no first order 
generally covariant theories  (i.e., based on a covariant Lagrangian) 
for symmetric tensor fields at all. Namely, since $\pi^{kl(m)}$ is symmetric 
in $kl$, it cannot be antisymmetric in $km$, except if $\pi^{kl(m)}$ is 
zero (that would be the case for a Lagrangian consisting only of a 
cosmological constant). It is not hard to derive the same conclusion for 
scalar densities (so-called dilaton fields) 
and for symmetric tensor densities.

On the other hand, in order to evaluate the stress-energy tensor, 
we are not really interested in $L^{mk}_{\ \ i}$, but rather in 
$L^{mk}_{\ \ i,m}$. Therefore, as we have already indicated in section 6,  
we can modify the second line  in (\ref{42}) and write instead 
$[\sqrt{-g}(-2g^{lm} \delta^k_i + 2 g^{lk} \delta^m_i)]_{,l}$. We denote  
by $\tilde L^{mk}_{\ \ i}$ 
the expression (\ref{42}) with the second line replaced in that way. 
This is now antisymmetric in $km$. Then, we have 
\begin{equation}\label{89} 
\tilde L^{mk}_{\ \ i}  - L^{mk}_{\ \ i} = [\sqrt{-g}(g^{lk} \delta^m_i 
- g^{km} \delta^l_i)]_{,l}. 
\end{equation}
Obviously, $\tilde L^{mk}_{\ \ i,m} = L^{mk}_{\ \ i,m}$. We can thus 
write $\tau^k_{\ i} = - \frac{1}{2} \tilde L^{mk}_{\ \ i,m}$, with 
$\tilde L^{mk}_{\ \ i} $ antisymmetric in $mk$. In view of (\ref{88}), 
we introduce modified momenta $\tilde \pi^{il(m)}$ by requiring 
\begin{equation} \label{90} 
\tilde L^{mk}_{\ \ i} = 4 \tilde \pi^{kl(m)} g_{li}, 
\end{equation}
which leads to 
\begin{equation} \label{91} 
\tilde \pi^{ki(m)} = \pi^{ki(m)} + (\sqrt{-g} g^{kl})_{,l} g^{mi} 
- (\sqrt{-g} g^{km})_{,l} g^{li}. 
\end{equation} 
Note that $\tilde \pi^{ki(m)}$ is no longer symmetric in $ki$. Next, 
we postulate canonical commutation relations between fields and 
modified momenta in the form (for spacelike separations) 
\begin{equation}\label{92} 
[g_{ik}(x), \tilde \pi^{ln(m)}(y)] =
\frac{1}{2}( \delta^l_i \delta^n_k+ \delta^l_k \delta^n_i)
\ \delta^m_{\sigma}(x-y). 
\end{equation}  
Only the symmetric part of $\tilde \pi^{ln(m)}$ does not commute with 
$g_{ik}$ and contains thus the physical momenta. 

The stress-energy tensor $\tau^i_{\ k} = 
(\partial \mathcal L/ \partial g_{lm,i})g_{lm,k} - \delta^i_k \mathcal L$ 
can now be written in the form 
\begin{equation}\label{93}
\tau^i_{\ k} = \pi^{lm(i)} g_{lm,k} - \delta^i_k \mathcal L 
= \tilde \pi^{lm(i)} g_{lm,k} -[(\sqrt{-g} g^{pl})_{,p} g^{im} - 
(\sqrt{-g} g^{li})_{,p} g^{pm}] g_{lm,k} 
- \delta^i_k \mathcal L.  
\end{equation}
We can now proceed as in the previous cases, i.e., we define the 
generators by 
\begin{equation} \label{94} 
P = \int \epsilon^k \tau^i_{\ k} \de \sigma_i, 
\end{equation}
as well as 
\begin{equation} \label{95} 
\tilde P = \int \epsilon^k \tilde \tau^i_{\ k} \de \sigma_i 
\end{equation} 
with $\tau^i_{\ k}$ as above and with 
\begin{equation}\label{96}
\tilde \tau^i_{\ k} = - \frac{1}{2} \tilde L^{im}_{\ k,m} 
= - 2 [\tilde \pi^{ml(i)} g_{kl}]_{,m}. 
\end{equation} 
Finally, for the third generator (see (\ref{83})), we have 
\begin{equation}\label{97} 
\hat P = \int [\epsilon \hat \tau^{mi}_{\ \ k}]_{,m} \de \sigma_i  
 =2 \int [\epsilon^k \tilde \pi^{ip(m)} ]_{,m} \de 
\sigma_i,  
\end{equation}
where $\hat \tau^{mi}_{\ \ k} = \frac{1}{2} \tilde L^{mi}_{\ \ k}$. 
The primary constraints are now expressed by  the fact that 
$\hat \tau^{mi}_{\ \ k}$  is antisymmetric, and thus $\hat P = 0$,   
while the secondary constraints are given by $P + \tilde  P = 0$. Thus, 
in a sense, with the modification of the canonical momenta, 
we have restored the primary constraints which have been lost because 
we started with a  non-covariant Lagrangian.  
 
The commutation relations of those generators with the metric lead  
consistently to the expressions (\ref{84}), (\ref{85}) and (\ref{85b}). 
Explicitely, if we choose again the hypersurface $t =const$, we have 
\begin{eqnarray} \label{99} 
[P, g_{ik}] &=& - g_{ik,m} \epsilon^m, \\ \label{100} 
[\tilde P, g_{ik}]&=& - \epsilon^m_{\ ,i} g_{mk} 
- \epsilon^m_{\ ,k} g_{mi} + (\epsilon^m g_{mk})_{,0} \delta^0_i 
+ (\epsilon^m g_{mi})_{,0} \delta^0_k, \\ \label{101}
[\hat P , g_{ik}] &=& -( g_{im} \epsilon^m)_{,0} \delta^0_k 
-( g_{km} \epsilon^m)_{,0} \delta^0_i. 
\end{eqnarray}
From this, we can see explicitely that action of the constraint 
$P + \tilde P=0$
on the spatial components, namely 
\begin{equation} \label{102}
[P + \tilde P, g_{\mu\nu} ]= 
- g_{\mu\nu, \lambda} \epsilon^{\lambda}
- \epsilon^{\lambda}_{\ ,\mu} g_{\kappa\nu} 
- \epsilon^{\lambda}_{\ ,\nu} g_{\kappa\mu},  
\end{equation}
where we have assumed that $\epsilon^0 = 0$, i.e., the transformation takes 
place on the hypersurface in question. As pointed out earlier, this 
is again the Lie derivative of the three dimensional metric. 
In Appendix B, we analyze  the question whether  
$g_{\mu\nu}$ can indeed be assumed to represent the propagating part of 
$g_{ik}$ in an appropriate gauge. 

In order to check whether  the introduction of the new momenta 
$\tilde \pi^{lm(i)}$ 
is fully consistent, one has to introduce a Hamiltonian 
$H = \int \tau^i_{\ k} n^k \de \sigma_i$ (see \cite{leclerc2}), 
express it in terms of 
the momentum $\tilde \pi^{lm} = \tilde \pi^{lm(i)}n_i$, 
i.e., eliminate the velocities 
$g_{lm,i}n^i$, and analyze the commutation relations of $H$ with the
momenta. This is a non-trivial task, in particular one will have 
to deal with ordering problems, see, e.g., \cite{zanelli}. It is 
however not unusual in the context of first order general 
relativity to  carry out modifications by hand, in order to achieve 
consistency between the constraints of the theory. For instance,  Dirac 
\cite{dirac2} chose to modify the Lagrangian by a surface term 
in a way that the momenta conjugated to $g_{0\mu}$ vanish weakly, which 
then represents a primary constraint.   

An alternative way, and probably a more  elegant one, 
is  to start from the covariant, 
second order Lagrangian. Let us briefly sketch how this could work. 
One defines the canonical momenta 
\begin{equation} \label{103}
\pi^{i} = \frac{\partial \mathcal L}{\partial \phi_{,i}} - (
\frac{\partial \mathcal L}{\partial \phi_{,k,i}})_{,k}, \ \ \ \  
p^{m(i)}= \frac{\partial \mathcal L}{\partial \phi_{,m,i}} 
\end{equation}
and writes the stress-energy tensor (\ref{53}) in the form 
\begin{equation} \label{104}
\tau^k_{\ i} = \pi^{(k)} \phi_{,i} + p^{m(k)} \psi_{m,i} - 
\delta^k_i \mathcal L, 
\end{equation}
where $\psi_m = \phi_{,m}$ plays the role of the canonical variable 
conjugate to $p^{m(i)}$. The identically conserved form of the 
stress-energy tensor from (\ref{54}) takes the simple form 
\begin{equation} \label{105}
\tau^k_{\ i} = - \frac{1}{2}\left[\pi^{(m)} (\sigma \phi)^k_{\ i} + 
p^{j(m)} (\sigma \psi_j)^k_{\ i}\right]_{,m}
 \end{equation}
where $(\sigma \psi_j)^k_{\ i} 
= 2 \delta^k_j \phi_{,i} + [(\sigma \phi)^k_{\ i}]_{,j}$, 
i.e., it acts correctly on $\psi_j$ in accordance with its total 
tensor structure, taking account of the additional vector index.  
Similarly, we express equations (\ref{55}) and (\ref{56}) in terms of 
the momenta and finally construct four charge operators, the 
first two expressing the constraints stemming from the 
equality between (\ref{104}) and (\ref{105}), and the other two 
expressing the vanishing of the expressions (\ref{55}) and (\ref{56}). 
For instance, in a metric theory, it can easily be seen from (\ref{56})
that  one set of (primary) constraints is related to the fact that no 
second time derivatives 
of the the components $g_{0i}$ are contained in the Lagrangian. We thus see 
that this apparent {\it coincidence} in general relativity is actually 
a necessary feature of generally covariant second order theories. 

The only non vanishing commutators in second order field theory are given by 
\begin{equation} 
[\phi(x), \pi^{(i)}(y)] = \delta^i_{\sigma}(x-y), \ \ 
[\psi_m(x), p^{k(i)}(y)] = \delta_m^k \delta^i_{\sigma}(x-y),
\end{equation}
where $x$ and $y$ are assumed to be separated by a spacelike separation. 

\section{Conclusions} 

We have analyzed Noether's theorem  localizing step by step 
the transformation group from global to local. The intermediate 
steps of linear (and eventually quadratic) transformations turned 
out to be of importance, not only in general relativity, where one 
usually prefers the  use of a first order 
Lagrangian which is covariant only under affine 
transformations, but also in theories based on a 
fully covariant Lagrangian. Indeed, the restrictions on the 
conserved current derived from each step of locality are directly 
related to the first class Hamiltonian constraints of the theory. 
The primary, secondary, etc., constraints  related to the invariance of 
the theory under a certain symmetry group, as well as 
the corresponding generators,  can be directly read off from the  
form of the  Noether current as it arises in each step. 

Concerning the stress-energy tensor, we have shown that it is 
identically conserved whenever the theory is generally covariant
and the momentum can thus be written in terms of two dimensional surface 
integrals. In first order theories, the symmetric 
Belinfante tensor was shown to be 
identically zero. In second order theories, this is not 
necessarily the case, but the generalization of 
the Belinfante formula to those theories does not seem to be unique.  
Special attention has been  paid to general relativity, which, in the 
first order approach,  shares many  features with 
theories based on a fully covariant 
Lagrangian, but at some points, modifications are necessary in order to 
restore the properties that have been lost as a result of the use of 
a non-covariant Lagrangian. 

\small

\appendix 

\subsubsection*{Appendix A: Covariant Hamiltonian formalism}

We recall the main features of the manifestly covariant formalism 
for field theory used in \cite{leclerc2}. 

Let $x \equiv x^i = (x^0, x^1, x^2, x^3)$  be 
spacetime  coordinates such that a hypersurface element can be written 
as  
\begin{equation}\label{app0} 
\de \sigma_i(x)  = \left( \begin{array}{c} 
\de x^1(\sigma) \de x^2(\sigma) \de x^3(\sigma) \\ 
 \de x^0(\sigma) \de x^2(\sigma) \de x^3(\sigma) \\  
\de  x^0(\sigma) \de x^1(\sigma) \de x^3(\sigma) \\ 
 \de x^0(\sigma) \de x^1(\sigma) \de x^2(\sigma)
\end{array}
\right), 
\end{equation}   
where  $\de x^i(\sigma) $ means that $\de x^i $ is restricted 
to some hypersurface $\sigma$ defined by $\Phi(x) = 0$. (E.g., for 
the hypersurface $x^0 = const$, we have $\de x^0 = 0$ and $\de \sigma_i(x) 
= \delta^0_i \de^3 x$.)

In the same coordinate system, we define 
\begin{equation} \label{app1} 
\delta^i (x-y) = 
\left( \begin{array}{c}
\delta_{x^0 y^0}\  \delta(x^1-y^1)\delta(x^2-y^2)\delta(x^3-y^3)\\
\delta_{x^1 y^1}\  \delta(x^0-y^0)\delta(x^2-y^2)\delta(x^3-y^3)\\
\delta_{x^2 y^2}\  \delta(x^0-y^0)\delta(x^1-y^1)\delta(x^3-y^3)\\
\delta_{x^3 y^3}\  \delta(x^0-y^0)\delta(x^1-y^1)\delta(x^2-y^2)
\end{array}
\right).
\end{equation}

The transformation behavior for  $\delta^i (x-y)$ 
under a coordinate change is  found 
from the known transformation behavior of $\de \sigma_i(x)$ 
($\sqrt{-g}\ \de \sigma_i $ is a vector)  
by requiring $\delta^i (x-y) \de \sigma_i $ to transform as  scalar under 
general coordinate transformations.  Thus, $\delta^i(x-y)$ transforms as 
vector density. 

Next, consider 
a spacelike hypersurface $\sigma$ defined by $\Phi(x) = 0$, with the  
 (timelike) normal 
 vector $n_i = \Phi_{,i}$. For convenience, $\Phi(x)$ can be chosen such 
that $n^2 \equiv n_{i}n_{k} g^{ik} = 1$. Then, we have 
\begin{equation} \label{app2}
\int_{\sigma} f(x) \delta^i(x-y)  \de \sigma_i(x) = f(y) 
\end{equation}
where the integration is carried out over the  hypersurface $\sigma$ 
containing 
the point $y$. For the specific hypersurface $t = t_0=const$, 
we find\footnote{We use both $x^0$ (and $y^0$ etc.) as well as  
$t$ for the time coordinate, while $t_0$ always refers to a constant.}, e.g., 
\begin{equation}  \label{app3} 
\int_{\sigma} 
f(x) \delta^i(x-y)  \de \sigma_i(x) = \int_{\sigma} 
 f(x^0, \vec x) 
\delta^{(3)}(\vec x - \vec y)
\ \delta_{x^0 y^0}\  \de^3 x 
= f(x^0, \vec y) \ \delta_{x^0 y^0}, 
\end{equation}
where $x^0$ is to be taken on the hypersurface in question, i.e., 
$x^0 = t_0$. Thus, if 
$y^0 = t_0$ (i.e., if $y$ lies  on the hypersurface $t=t_0$), 
the result is simply 
$f(y)$, while else, we find zero. Thus,  $\delta^i (x-y) $ can be seen as 
 covariant generalization of the three dimensional delta function.

Two useful relations are the following (see \cite{leclerc2}) 
\begin{equation}\label{app4}
 \int_{\sigma} f_{,i} \de \sigma_k = \int_{\sigma} f_{,k} \de \sigma_i. 
\end{equation}
and 
\begin{equation} \label{app5} 
\int_{\sigma} f(x) n_i \de \sigma_k (x) = \int_{\sigma} 
f(x) n_k \de \sigma_i(x). 
\end{equation}
Note that in general, $n_i = n_i(x)$, 
but we will omit the argument whenever there is no 
danger of confusion. In order for the above relations to hold, an appropriate 
asymptotical behavior of $f$ has to be assumed, such that two dimensional 
 integrals over the boundary of $\sigma$ can be omitted.

In particular, we have 
\begin{eqnarray} \label{app6}
\int_{\sigma} 
f(x) \delta^i(x-y)  \de \sigma_i &=& 
\int_{\sigma} f(x)  \delta^i (x-y) n^k n_k \de \sigma_i  
\nonumber \\
&=& \int_{\sigma}  f(x) \delta^i (x-y)  n_i n^k \de \sigma_k  
\end{eqnarray} 
which is equal to $f(y)$ if $y$ lies on the hypersurface. (We take the 
convention that all quantities whose arguments are not written explicitely 
are to be taken at the point $x$.) 
Let us introduce 
the following definitions 
\begin{eqnarray} \label{app7}
\de \sigma &=& n^i \de \sigma_i \\ \label{app8}
\delta_{\sigma} (x-y) &=& \delta^i (x-y) n_i \\ \label{app9}
\delta_{\sigma}^i (x-y) &=& \delta_{\sigma}(x-y) n^i 
= \delta^m (x-y) n_m n^i.  
\end{eqnarray}
We can thus write 
\begin{equation} \label{app10}
\int_{\sigma}  f(x) \delta^i (x-y) \de \sigma_i (x) = \int_{\sigma}  
f(x) \delta^i_{\sigma} (x-y) 
\de \sigma_i(x) = \int_{\sigma} f(x)\delta_{\sigma}(x-y) \de \sigma = f(y) 
\end{equation} 
where for the last relation, 
it is assumed that $y$ lies on the hypersurface. Moreover, we have
 $\delta^i (x-y) n_i = \delta_{\sigma}^i (x-y)n_i $. Nevertheless, one should 
not confuse  $\delta^i (x-y)$ which is given explicitely by (\ref{app1}), 
with $\delta^i_{\sigma} (x-y)$, which is defined with respect to a specific 
hypersurface. In particular, for $t = t_0$, we have $\delta^i_{\sigma}(x-y) = 
\delta_{x^0 y^0}\  \delta^{(3)}(\vec x - \vec y) g^{0i}/g^{00}$. (The factor 
involving the metric components stems from the normalization $n^2 = 1$, 
which for $t=t_0$ (and thus $n_i = (n_0, 0, 0, 0)$ ) 
leads to $n_0 = 1/\sqrt{g^{00}}$ and $n^i = g^{0i} n_0$.) In particular, 
in flat spacetime, we see that $\delta^i_{\sigma}(x-y)$ has only 
one non-vanishing component, in contrast to (\ref{app1}). 

In order to derive the commutation relations with the charges 
we encounter in section 8, the following relation is useful 
\begin{equation} \label{app11} 
\int_{\sigma} f(x) [\delta^i_{\sigma}(x-y)]_{,l} \de \sigma_i(x) 
= - f_{,l}(y) + f_{,i}(y) n^i(y) n_l + \int_{\sigma} f(x) 
[\delta^i_{\sigma} (x-y)]_{,i} \de \sigma_l(x), 
\end{equation} 
which is easily derived with the help of (\ref{app4}) and (\ref{app5}). 
The last term in (\ref{app11}) cannot be simplified without 
specifying a hypersurface. (Note that $[\delta^i(x-y)]_{,i} = 0$, 
but $[\delta_{\sigma}^i(x-y)]_{,i} \neq  0$.) For the hypersurface 
$t =const$, we find by partial integration $-f_{,\mu} (g^{0\mu}/g^{00}) 
\delta^0_l$, and thus 
\begin{equation} \label{app12} 
\int_{\sigma} f(x) [\delta^i_{\sigma}(x-y)]_{,l} \de \sigma_i(x) 
= - f_{,l} + f_{,0} \delta^0_l.  
\end{equation}
With the help of those relations, it is an easy task to 
evaluate the commutators (\ref{75})-(\ref{77}) 
and  (\ref{84})-(\ref{86}).

\subsubsection*{Appendix B: Propagating 
fields in linearized general relativity}

We consider a perturbation $h_{ik}$ on a flat background $g_{ik} = 
\eta_{ik} + h_{ik}$. The field equations of general relativity, 
$\sqrt{-g} G_{ik} = \mathcal T_{ik}$ to first order in $h_{ik}$ 
have the form  
\begin{equation} \label{3aap}
\frac{1}{2}[- \Box \psi_{ik} + \psi^l_{\ i,k,l}+ \psi^l_{\ k,i,l}
- \eta_{ik} \psi^{lm}_{\ \ ,l,m}] = \mathcal T_{ik},  
\end{equation}
where $\psi_{ik} = h_{ik} - \frac{1}{2} h$, and $h = \eta^{ik} h_{ik}$. 
In particular, to this order, we have the conservation law 
$\mathcal T^{ik}_{ \ \ ,i} = 0$. Further, we have the 
gauge freedom of the linearized theory, $\delta h_{ik} = \xi_{i,k} + 
\xi_{k,i}$, where $\xi^i $ is assumed to be of the same order as $h_{ik}$. 
All indices are raised and lowered with $\eta_{ik}$. 

Our intention is to look for a Coulomb type gauge choice, where the 
non-propagating fields can be eliminated by solving the field equations 
(Gauss' law in electromagnetism) and the remaining fields satisfy a 
conventional wave equation. Note that, with the conventional gauge 
choice $\psi^{ik}_{\ \ ,i} = 0$, the field equations take the form 
$\mathcal T_{ik}  = - \frac{1}{2} \Box \psi_{ik}$, which is easy to 
solve, but it does not tell us anything on the number of propagating 
fields (there are still  6 independent fields). 

A more convenient gauge choice can be read off the field equations after 
a 3+1 split,  
\begin{eqnarray} \label{7ap}
\mathcal 
T_{00} &=& \frac{1}{2}( \Delta \psi_{00} - \psi^{\mu \nu}_{\ \ ,\mu, 
\nu}) \\ \label{8ap}
\mathcal 
T_{0 \mu} &=& \frac{1}{2}( \Delta \psi_{0 \mu} + 
\psi^{\nu}_{\ \mu,0,\nu}  +\psi^{\nu}_{\ 0,\mu,\nu}  
+ \psi^0_{\ 0,\mu,0}) \\ \label{9ap}
\mathcal 
T_{\mu \nu} &=& \frac{1}{2}(-\Box \psi_{\mu \nu} + \psi_{0\mu,0,
\nu} + \psi_{0\nu,0,\mu}+ \psi^{\lambda}_{\ \nu,\lambda,\mu}
+ \psi^{\lambda}_{\ \mu,\lambda,\nu}) 
- \frac{1}{2} \eta_{\mu\nu}(\psi^{00}_{\ \ ,0,0}+ \psi^{\lambda\kappa}_{\
  \ ,\lambda,\kappa}+ 2 \psi^{0\lambda}_{\ \ ,0,\lambda}),
\end{eqnarray}
where $\Delta = -\partial^{\mu} \partial_{\mu}$ and $\Box = 
\partial^2_0 - \Delta$.  We choose the following gauge conditions  
\begin{equation} \label{10ap}
\psi^{\mu \nu}_{\ \ ,\nu} = 0, \ \ \psi^{0m}_{\ \ ,m} = 0.
\end{equation}
The first condition on the three four-vectors $\psi^{\mu m}$ 
is analogous  to the Coulomb gauge in Maxwell theory, while 
the second one can be seen as a Lorentz gauge on the 
four-vector $\psi^{0m}$. To show that it is indeed possible to impose 
such conditions on $\psi^{ik}$, consider the variations under 
gauge transformations of the quantities in question. We find  
\begin{equation}\label{11ap}
\psi^{0m}_{\ \ ,m} \rightarrow \psi^{0m}_{\ \ ,m} + \Box \xi^0, \ \ 
\psi^{\mu\nu}_{\ \ ,\nu}
\rightarrow \psi^{\mu\nu}_{\ \ ,\nu} + \xi^{0\ \ ,\mu}_{\ ,0}
- \xi^{\mu,\nu}_{\ \ ,\nu}.
\end{equation}
Therefore, we  perform first a transformation with 
$\xi^0$ satisfying $\Box \xi^0 = 
- \psi^{0m}_{\ \ ,m}$ (and with $\xi^{\mu}=0$), and then a transformation 
with $\xi^{\mu}$ satisfying 
$\Delta \xi^{\mu} = - \psi^{\mu\nu}_{\ \ ,\nu}$ (and 
with $\xi^0 = 0$).  Both 
transformations exists generally, and since the second transformation 
does not modify the result of the first, the argument is complete. 
Note that it is not possible to require 
$\psi^{\mu\nu}_{\ \ ,\nu}= 0$ together with 
$\psi^{0\mu}_{\ \ ,\mu} = 0$ instead of (\ref{10ap}).

The field equations take the simplified form 
\begin{eqnarray} \label{12ap}
\mathcal T_{00} &=& \frac{1}{2} \Delta \psi_{00} 
 \\ \label{13ap}
\mathcal T_{0 \mu} &=& \frac{1}{2} \Delta \psi_{0 \mu}  
 \\ \label{14ap}
\mathcal T_{\mu \nu} &=& - \frac{1}{2} \Box \psi_{\mu \nu} 
+\frac{1}{2}( \psi_{0\mu,0,\nu} + \psi_{0\nu,0,\mu}
-  \eta_{\mu\nu} \psi^{0\lambda}_{\ \ ,0,\lambda}). 
\end{eqnarray}
The residual gauge freedom is now restricted to 
\begin{equation}\label{15ap}
\Box \xi^0 = 0, \ \ \ \ \xi^{0,\mu}_{\ \ ,0}+ \Delta \xi^{\mu} = 0.
\end{equation}
Similar as in electrodynamics, we can eliminate the 
non-propagating modes by solving (\ref{12ap}) and (\ref{13ap}). We have 
\begin{eqnarray} \label{16ap}
\psi_{00}(t,x)  &=& -\frac{1}{2\pi} \int 
\frac{\mathcal T_{00}(t,x')}{|x - x'|} 
\de^3 x', \\
\label{17ap}
\psi_{0\mu} (t,x) &=& -\frac{1}{2\pi} \int \frac{\mathcal T_{0\mu}(t,x')}
{|x - x'|} \de^3 x'.
\end{eqnarray}
It is an easy task to verify that this solution satisfies indeed 
the constraint $\psi^{0m}_{\ \ ,m} = 0$.
They are Coulomb type solutions, and for slowly moving 
matter distributions, the important contribution is given by (\ref{16ap}), 
corresponding to the Newtonian potential.  

To some extend, we have obtained a system quite similar to  
 electrodynamics in the Coulomb gauge. As was to be expected, to 
the charge density correspond the four momentum density 
components $\mathcal T_{i 0}$, and the corresponding fields $\psi_{i 0}$ 
can be eliminated by a Gauss type law. Also, 
just like $ A^{\mu}_{\ ,\mu} = 0$, we have a remaining, symmetric 
transverse tensor satisfying $\psi^{\mu \nu}_{ \ \ ,\mu} = 0$,  
    containing the propagating modes. 
Further, the non-propagating modes appear as an additional source 
to the tensor $\psi_{\mu\nu}$ in  (\ref{14ap}). This too is in 
straight analogy to  the  Maxwell equation 
$ \Box A^{\mu} = - j^{\mu} + A^{0,\mu}_{\ \ ,0}$ in the 
Coulomb gauge. 

To simplify equation (\ref{14ap}), we observe that at large distances 
from the matter distribution (far zone), we have approximately 
$|x-x'| \approx |x|$, such that the four equations (\ref{16ap}) and 
(\ref{17ap}) can be approximated by 
\begin{equation} \label{18ap}
\psi_{0i} (t,x) = - \frac{1}{2 \pi}\ \frac{1}{|x|} \int 
\mathcal T_{0i}(t,x') \de^3 x'. 
\end{equation}
On the other hand, we have in the linear theory 
the relation $\mathcal T^{ik}_{\ \ ,k} = 0$ which leads 
upon integration to the 
conservation law $ (\de / \de t) \int \mathcal T^{i0} \de^3 x = 0 $. 
In particular, from  (\ref{18ap}), we thus have $\psi_{0i,0} = 0$, 
and therefore equation (\ref{14ap}) reduces to 
\begin{equation} \label{19ap}
\Box \psi_{\alpha \beta} = - 2 \mathcal T_{\alpha \beta}.  
\end{equation}
We also retain that from  $\psi_{0i,0}= 0$, together with our gauge 
conditions (\ref{10ap}), we find 
\begin{equation} \label{new}
\psi^{il}_{\ \ ,l} = 0\ \ \text{and}\ \  
\psi^{\alpha\beta}_{\ \ ,\beta} = 0\ \ \text{for}\ \ |x| >> \ell, 
\end{equation}
where $\ell$ characterizes the dimensions of the matter distribution. 
Therefore, the conventional Lorentz type gauge condition holds 
again in this limit. Together with (\ref{19ap}), we see that our 
field equations, as far as $\psi_{\mu\nu} $ is concerned, are completely 
equivalent to those of the  conventional approach based on 
the gauge $\psi^{ik}_{\ \ ,k}$. Nevertheless, we are 
one step ahead, because we did  not simply \textit{forget}  the 
components $\psi_{\mu 0}, \psi_{00}$, but we have eliminated them 
properly via a Gauss type law.  

The above approach  is probably the closest one can get to 
the Coulomb analogy of Maxwell's theory. There is only one disturbing 
point: The field $\psi^{\mu\nu}$ satisfying the transversality 
condition $\psi^{\mu\nu}_{\ \ ,\mu} = 0$ contains still 
three independent degrees of freedom, which is one more than those needed 
for the description of the massless spin 2 field. It is not hard to 
see that for plane wave solutions, the residual gauge freedom (\ref{15ap}) 
can be used to make $\psi^{\mu\nu}$ traceless, thus excluding the 
existence of an additional scalar field. It should therefore be possible 
to eliminate a fifth component of $\psi^{ik}$ either by 
 an additional Gauss type law, e.g., 
the trace of $\psi^{ik}$ or of $\psi^{\mu\nu}$ and relate it 
to a fifth charge (e.g., the trace of $\mathcal T^{ik}$ or of 
$\mathcal T^{\mu\nu}$), or simply 
to impose an additional condition of $\psi^{\mu\nu}$ to reduce the 
degrees of freedom to two.  However, 
with the gauge freedom (\ref{15ap}), this does not seem to be  possible. 
Neither would  
the appearance of a fifth charge be physically satisfying. We therefore 
suggest that there exists a better gauge choice, such that upon eliminating 
four components by a Gauss type law and imposing four conditions on 
the remaining components, the gauge is completely fixed, and the degrees of 
freedom are reduced to two. We do not know whether such a gauge has been 
proposed in literature. 

In any case, we see that the propagating modes (in a perturbative 
approach) can be assumed to 
be contained in the spatial components of $\psi_{ik}$, 
which is (up to a constant) the linear approximation 
of $\frac{1}{\sqrt{-g}}g_{ik}$,  the tensor with inverse 
determinant of $g_{ik}$. It is now an easy matter to evaluate the 
action of the generator $P + \tilde P$ on this field, 
using (\ref{84}) and (\ref{85}). We find 
\begin{equation}
[P+ \tilde P, \frac{1}{\sqrt{-g}} g_{\mu\nu}] = 
- (\frac{1}{\sqrt{-g}}g_{\mu\nu})_{,\lambda } \epsilon^{\lambda} 
- (\frac{1}{\sqrt{-g}} g_{\mu\lambda}) \epsilon^{\lambda}_{\ ,\nu} 
- (\frac{1}{\sqrt{-g}} g_{\nu\lambda}) \epsilon^{\lambda}_{\ ,\mu} 
  + (\frac{1}{\sqrt{-g}} g_{\mu\nu})\epsilon^{\lambda}_{\ ,\lambda}, 
\end{equation}
where we have assumed that $\epsilon^0 = 0$. This is the Lie 
derivative of the pseudotensor $(1/\sqrt{-g}) g_{\mu\nu}$, confirming 
our assumption that the secondary constraint $P  + \tilde P = 0$ 
generates  coordinate transformations (on the hypersurface $t= const$) 
on the propagating components of the field, in complete analogy to the 
case of non-abelian gauge theory. 

Finally, we note that the gauge choice (\ref{10ap}) is the linearized 
version of $(\sqrt{-g} g^{0i})_{,i}= 0 $ and $(\sqrt{-g} g^{\mu\nu})_{,\mu}
= 0$.

\end{document}